\documentclass[aip,jcp,preprint,notitlepage,amsmath,amssymb]{revtex4-2}
\usepackage[T1]{fontenc}
\usepackage{lmodern}
\usepackage[english]{babel}
\usepackage{bm}
\usepackage{graphicx}
\usepackage{microtype}
\usepackage{setspace}
\usepackage{hyperref}
\AtBeginDocument{\singlespacing\raggedbottom}
\newcommand{\angstrom}{\mbox{\normalfont\AA}}
\newcommand{\intk}{\int\!\frac{d^3\mathbf{k}}{(2\pi)^3}}
\hypersetup{colorlinks=true,citecolor=blue,linkcolor=blue,urlcolor=blue}

\begin{document}

\author{Yury A. Budkov}
\email[]{ybudkov@hse.ru}
\affiliation{Laboratory of Computational Physics, HSE University, Tallinskaya st. 34, 123458 Moscow, Russia}
\affiliation{School of Applied Mathematics, HSE University, Tallinskaya st. 34, 123458 Moscow, Russia}
\affiliation{Laboratory of Multiscale Modeling of Molecular Systems, G.A. Krestov Institute of Solution Chemistry of the Russian Academy of Sciences, 153045, Akademicheskaya st. 1, Ivanovo, Russia}
\author{Nikolai N. Kalikin}
\affiliation{Laboratory of Computational Physics, HSE University, Tallinskaya st. 34, 123458 Moscow, Russia}
\affiliation{Laboratory of Multiscale Modeling of Molecular Systems, G.A. Krestov Institute of Solution Chemistry of the Russian Academy of Sciences, 153045, Akademicheskaya st. 1, Ivanovo, Russia}

\title{Hall effect in electrolyte solutions: Self-consistent Debye--H\"uckel--Onsager theory}
\date{20 November 2026}

\begin{abstract}
A self-consistent statistical-mechanical theory of the Hall effect in electrolyte solutions is developed by extending the self-consistent Debye--H\"uckel--Onsager (SCDHO) framework to crossed electric and magnetic fields.
The theory incorporates non-local ionic charge distributions through Slater-type form factors and ion-specific coefficients describing the effective Lorentz response, with experimental input for protons and hydroxide ions.
Within the random-phase approximation, the Hall conductivity is expressed as mean-field, correlation, and electrophoretic contributions. Both interaction channels include the Lorentz rotation of the complete electric-field-induced ionic velocity.
In the point-charge limit with hydrodynamics suppressed, the relaxation correction agrees with independent odd stochastic density functional theory and the relaxation part of Friedman's classical result.
For 22 dilute aqueous electrolytes, the mean absolute relative deviations from measured Hall constants and Hall numbers are about 17\%.
At finite concentrations, the theory predicts a nonmonotonic Hall number for NaBPh$_4$ and sign changes of the electrophoretic correction for NaI and HCl. Correlation effects dominate for HCl and LiOH, whereas the two interaction channels become comparable for KOH and aqueous imidazolium salts.
The results clarify how ionic mobilities, effective Hall coefficients, charge distributions, and solvent hydrodynamics determine electrolyte magnetotransport.
\end{abstract}

\maketitle

\noindent\textbf{Revision note.} This version incorporates corrections to the Hall-current decomposition and numerical Hall results of the published article~\cite{budkov2026hall}. The Lorentz rotations of the ordinary relaxation and electrophoretic currents are included throughout. All nine Hall-dependent figures have been replaced, and incorrectly reproduced HNO$_3$ Hall measurements have been removed. The ordinary conductivity calculations and fitted parameters are unchanged. Equation numbers refer to this revised version.

\section{Introduction}
The Hall effect in electrolyte solutions constitutes a multifaceted interdisciplinary phenomenon, situated at the confluence of electrochemistry, hydrodynamics, and magnetism. While in solid-state systems this effect has long served as a reliable tool for characterizing charge carriers, its manifestation in liquid media involves fundamentally different physical mechanisms. In the presence of a magnetic field $\mathbf{B}$, an ion of charge $q$ moving with velocity $\mathbf{v}$ experiences the Lorentz force $\mathbf{F} = q[\mathbf{v} \times \mathbf{B}]$, which induces a transverse migration of ionic fluxes. As a result, a transverse potential difference -- the Hall voltage $V_H$ -- develops across the electrochemical cell, scaling with the applied fields and the concentration of charge carriers.

A major step toward a rigorous statistical-mechanical treatment of the Hall effect in electrolytes was taken by Friedman, who developed \cite{friedman1965calculation} the first microscopic theory of the Hall effect in ionic solutions within the 'brownon' model. Using a cluster expansion~\cite{friedman1965electrophoretic,friedman1965relaxation} of the conductivity, he derived explicit expressions for the ideal and limiting-law contributions to the Hall conductance and demonstrated that, although weak, the effect should be experimentally accessible for highly mobile ions. This approach was later substantially refined by Sung and Friedman~\cite{sung1987microscopic}, who employed the Mori projection-operator formalism to obtain an exact time-correlation function formulation of the ionic Hall number. Their analysis isolated the ponderomotive force contribution arising from the coupling between the vorticity and spin fields in the solvent and clarified how ionic Hall number depends both on the hydrodynamic boundary conditions at the ion surface and on the dielectric properties of the medium. Despite this conceptual progress, both formulations inherit the considerable technical complexity of the underlying cluster-expansion machinery: the number of relevant diagrams grows rapidly with the inclusion of higher-order concentration terms or multi-component mixtures, and the resulting expressions for the Hall conductivity quickly become analytically unwieldy. Moreover, the fundamentally perturbative nature of the cluster expansion limits its ability to capture the non-perturbative correlation effects that become essential at elevated concentrations. A non-perturbative theoretical framework, capable of describing the Hall effect over a wide concentration range without an explosion of diagrammatic complexity, would therefore be highly desirable.

A quantitative theoretical description of the Hall effect in electrolytes is complicated by the interaction of the ions with the surrounding solvent. Hubbard and Wolynes demonstrated that electrohydrodynamic forces arising from the polarization of the medium by a moving ion reduce the effective Lorentz force by approximately $30\%$  compared with its vacuum value \cite{hubbard1981electrohydrodynamic}. For slip boundary conditions at the ion-solvent interface, Kroh and Felderhof found a reduction of about $25\%$ \cite{kroh1987hall}. This effect, often captured by the dielectric friction coefficient $\alpha = 1 - \frac{7}{24}\frac{\varepsilon_r - 1}{\varepsilon_r}$ (where $\varepsilon_r$ is the relative permittivity of the solvent), leads to a systematic deviation of the Hall number from unity and must be incorporated into any quantitative model of electrolyte conductivity in magnetic fields.

Experimental observation of these effects poses considerable technical challenges. Typical Hall voltages in solutions lie in the range $0.1$--$1$~$\mu\text{V}$, whereas parasitic signals arising from imperfect electrode alignment and electromagnetic induction can exceed the sought-after signal by orders of magnitude. To overcome this problem, Meton and Gerard employed differential alternating-current techniques with cross-modulation \cite{meton1976hall}, while Bellissent et al. developed symmetrical measurement circuits based on quartz cells and careful spatial orientation~\cite{bellissent1971eperimental}.  These approaches enabled the first reliable detection of the Hall signal in extremely dilute solutions. Subsequent measurements by Gerard {\it et al.}~\cite{gerard1990hall} on aqueous HCl, H$_2$SO$_4$, and CH$_3$COOH solutions confirmed the robustness of the AC cross-modulation technique and yielded an ionic Hall number for the proton $\approx 0.1$, almost an order of magnitude smaller than the predictions of the existing continuum theories. This striking discrepancy underscores the role of the molecular structure of the solvent and the anomalous Grotthuss and hopping transport of the proton, highlighting the need for a microscopic theory that can separately address the Hall effect for structure-diffusing and hydrodynamic ions.

Magnetic fields influence not only bulk ion transport but also the conductivity of the electrolyte. Mohanta and Fahidy proposed a simplified framework that introduces the concept of an apparent ion-cloud radius for electrolytes subjected to crossed electric and magnetic fields~\cite{mohanta1976effect}. Their theory offers a rapid estimate of the ratio between the Hall conductivity and the ordinary conductivity and reveals that the effect becomes appreciable once the ionic cyclotron frequency $\omega_c = qB/m$ approaches the collision frequency, a condition that necessitates sufficiently strong fields.

Of particular interest are the effects observed at the electrode/electrolyte interface. Dunne and Coey experimentally established that even a moderate in-plane magnetic field of $B \approx 0.5~\text{T}$ can alter the double-layer capacitance and charge-transfer resistance by up to $50\%$ \cite{dunne2019influence}. These modifications are attributed to Maxwell stresses acting on paramagnetic species in solution, which induce a shift of the outer Helmholtz plane on the order of $0.25~\text{nm}$. The magnetic field thus becomes a tool for the fine-tuning of interfacial processes, including nucleation and wetting. It is important to note, however, that the Bohr--van Leeuwen theorem~\cite{kaufman2025lectures} guarantees that the equilibrium structure of the electrolyte remains unchanged by the magnetic field, a fact confirmed by molecular simulations~\cite{mouhat2013charge}. Consequently, the observed magnetic-field effects must be of nonequilibrium or pure quantum origin.

Practical applications of the Hall effect in electrolytes span several areas. Wright and van der Beken demonstrated the feasibility of exploiting the Hall principle for measuring electrolyte flow rates, including biomedical systems for monitoring blood flow~\cite{wright1972hall}. The effect is also considered a promising analytical method for determining ionic properties and concentrations and for the monitoring of industrial fluids and Hall biosensors~\cite{tamanaha2003magnetic,izci2019constructing}, although the complexity of precise small-signal measurements remains the principal obstacle~\cite{johnstone2008there}.

Although the Hall effect in electrolyte solutions has never been a mainstream topic of research, the emerging technological demands of nanofluidics\cite{aluru2023fluids,alizadeh2021electroosmotic,zhang2021nanofluidics}, iontronics~\cite{chun2015iontronics,qian2023bionic,kamsma2024brain}, and magneto-electrochemistry~\cite{salinas2023unconventional,sebastian2025magnetic} are creating a renewed need for a quantitative understanding of magneto-electrochemical transport. Recent progress in experimental techniques and theoretical modeling is gradually reviving interest in this phenomenon, transforming it from a barely detectable laboratory curiosity into a potentially useful tool for physical chemistry, electrochemistry, and engineering.

The theory of the Hall effect in electrolyte solutions can be developed similarly to the theory of electrical conductivity. 
In recent years, the theory of electrical conductivity in electrolyte solutions has undergone a period of vigorous development at widely varying levels of detail - from phenomenological generalizations of the classical Debye-H\"uckel-Onsager picture to rigorous microscopic approaches based on stochastic density functional theory and integral equations for correlation functions. Among the key directions, one should highlight stochastic density functional theory \cite{demery2016conductivity,avni2020charge,avni2022conductance}, which for the first time derived closed-form expressions for the relaxation and electrophoretic contributions to conductivity from first principles of nonequilibrium statistical mechanics; models based on the mean spherical approximation \cite{bernard1992conductance,bernard2023analytical}, which successfully describe concentrated solutions; and mean-field theory in which the electrophoretic effect is assigned the dominant role~\cite{vinogradova2023electrophoresis,vinogradova2025conductivity}. These works have substantially extended the predictive power of the theory, yet in most of them ions are still treated as point-like or spherically-symmetric objects with a localized charge, which limits the applicability of the models in the high-concentration regime where interionic distances become comparable to the dimensions of the solvation shells. An important step towards overcoming this limitation has been made in the recently developed self-consistent Debye-H\"uckel-Onsager (SCDHO) theory, proposed by us~\cite{budkov2026conductivity}. This theory explicitly accounts for the non-local charge distribution of solvated ions, using Slater-type charge form factors, which regularizes the Coulomb potential at short distances and eliminates the non-physical singularities present in models with point-like charges. The SCDHO theory allows for a natural division of total conductivity into three components: mean-field, correlation, and electrophoretic. It demonstrates high accuracy in describing aqueous and non-aqueous systems over a wide range of concentrations and temperatures. In addition, SCDHO introduces an effective charge-smearing length for each ion type. This parameter characterizes the width of the electrostatic form factor entering the coarse-grained Coulomb interaction. It should not be identified directly with either a bare crystallographic radius or a hydrodynamic solvated radius; rather, it represents the spatial scale over which the ionic charge distribution is resolved in the present electrostatic model.

The SCDHO theory provides a natural framework for describing the Hall effect in electrolyte solutions. Its formulation draws on precisely the same fundamental ingredients as the theory of electrical conductivity -- the two-particle continuity equation, the local chemical potentials from classical density functional theory, the Ornstein-Zernike relations, and the linearized Navier-Stokes equation for the solvent. To incorporate the magnetic field, the only additional terms required are the Lorentz force acting on the ions and the magnetic contribution to the force density in the Navier-Stokes equation, both renormalized to account for dielectric friction. This structural kinship means that the regularization of the Coulomb interaction at short distances, already achieved in SCDHO through Slater-type charge form factors, can be carried over to the Hall problem without introducing a fundamentally new theoretical apparatus. In the present paper, we implement this program and develop a self-consistent theory of the Hall effect in electrolyte solutions grounded in the SCDHO formalism.

\section{Theory}

\subsection{General equations}
Following the self-consistent Debye-H\"uckel-Onsager (SCDHO) theory~\cite{budkov2026conductivity}, we consider a sufficiently large and homogeneous system containing, in the most general case, $s$ species of ions, indexed by $a=1,\dots,s$ with mean concentrations $\bar{c}_a$. The ions carry total charges $q_a$. The solvent is treated as a continuous, incompressible fluid with shear viscosity $\eta$ and dielectric permittivity $\varepsilon$. An external, constant, homogeneous electric field $\mathbf{E}$ and magnetic field $\mathbf{B}$ are applied, driving the ions into a steady state of conduction. The derivation below is organized in four steps: the two-particle transport equation, the linearized chemical potentials, the hydrodynamic response of the solvent, and the resulting decomposition of the Hall current.

\subsubsection{Two-particle transport equation.}

Following Onsager and Fuoss~\cite{onsager2002irreversible,fuoss1978review}, we adopt a two-particle correlation approximation for the continuity equation within the configuration space of ions in an electrolyte solution. Namely, in the configuration space of two ions, $a$ and $b$, in a steady state the following continuity equation for the probability flows holds~\cite{pitaevskii2012physical}
\begin{equation}
\label{continuous}
\nabla_a\cdot \mathbf{j}_{ab}(\mathbf{r}_a,\mathbf{r}_b) +\nabla_{b}\cdot\mathbf{j}_{ba}(\mathbf{r}_b,\mathbf{r}_a)=0,
\end{equation}
where the operators $\nabla_a$ and $\nabla_b$ mean that differentiation is performed over the ion positions $\mathbf{r}_a$ and $\mathbf{r}_b$, respectively;
\begin{equation}
\label{flux}
\mathbf{j}_{ab}(\mathbf{r}_a,\mathbf{r}_b)=g_{ab}(\mathbf{r}_a,\mathbf{r}_b)\mathbf{v}_{ab}(\mathbf{r}_a,\mathbf{r}_b),\quad a\leftrightarrow b
\end{equation}
is the probability flow for ion $a$ at point $\mathbf{r}_a$ when ion $b$ is fixed at point $b$ and {\sl vice versa} with $g_{ab}(\mathbf{r}_a,\mathbf{r}_b)$ being the steady-state pair correlation function of ions $a$ and $b$. The total velocity of ions $a$ around ion $b$ is
\begin{equation}
\label{velocity}
\mathbf{v}_{ab}(\mathbf{r}_a,\mathbf{r}_b)=m_a \mathbf{f}_{a}(\mathbf{r}_a,\mathbf{r}_b)-m_a\nabla_a \mu_{a}(\mathbf{r}_a,\mathbf{r}_b)+\mathbf{u}_b(\mathbf{r}_a,\mathbf{r}_b),\quad a\leftrightarrow b,
\end{equation}
where
\begin{equation}
\mathbf{f}_{a}(\mathbf{r}_a,\mathbf{r}_b)=q_a\mathbf{E} + \beta_a q_a[\mathbf{v}_{ab}(\mathbf{r}_a,\mathbf{r}_b)\times \mathbf{B}]
\end{equation}
is the total force acting on the ion of kind $a$, with the first term being the electrostatic force and the second term on the right-hand side the effective Lorentz force. The coefficient $\beta_a$ accounts for the reduction of the Lorentz force due to microscopic solvent effects. For ions whose transport obeys classical Brownian dynamics, $\beta_a$ is given by the dielectric friction coefficient

\begin{equation}
\label{beta_a_det}
\beta_a \equiv \alpha = 1 - \frac{7}{24}\frac{\varepsilon_r-1}{\varepsilon_r},
\end{equation}
where $\varepsilon_r=\varepsilon/\varepsilon_0$ is the relative permittivity and $\varepsilon_0$ is the vacuum permittivity, as derived by Hubbard and Wolynes~\cite{hubbard1981electrohydrodynamic} and Kroh and Felderhof~\cite{kroh1987hall}. For protons and hydroxide ions in water, the charge migrates via the Grotthuss hopping mechanism rather than by Stokesian drift. As a result, the effective Lorentz force is strongly suppressed, and $\beta_{\mathrm{H}^+}$ and $\beta_{\mathrm{OH}^-}$ are significantly smaller than $\alpha$, with values around $0.1$ and $0.2$ inferred from the Hall measurements of Meton and Gerard~\cite{meton1976hall,gerard1990hall}. The numerical comparisons below use ion-specific effective coefficients; their adopted values are listed in Table~\ref{input_parameters}.

In the following, $m_a$ is the mobility of ion $a$, $q_a$ is its total electric charge, and $\mathbf{u}_b(\mathbf{r}_a,\mathbf{r}_b)$ is the average solvent velocity field at point $\mathbf{r}_a$ around ion $b$ placed at point $\mathbf{r}_b$.

\subsubsection{Linearized chemical potentials and pair correlations.}

The local chemical potential $\mu_{a}(\mathbf{r}_a,\mathbf{r}_b)$ is a functional of the local conditional concentrations of ionic species, $c_a(\mathbf{r}_a,\mathbf{r}_b)=\bar{c}_a g_{ab}(\mathbf{r}_a,\mathbf{r}_b)$. The total nonequilibrium chemical potential of ions $a$ can be written as a sum of ideal-gas and excess contributions,
\begin{equation}
\label{chem_pot_full}
\mu_{a}(\mathbf{r}_a,\mathbf{r}_b)=k_B T\ln\bigl(\lambda_a^3\bar{c}_a g_{ab}(\mathbf{r}_a,\mathbf{r}_b)\bigr)+\mu^{(\text{ex})}_{a}(\mathbf{r}_a,\mathbf{r}_b),\quad a\leftrightarrow b,
\end{equation}
where $k_B$ is the Boltzmann constant, $T$ is the temperature, and $\lambda_a$ is the de Broglie thermal wavelength of ion $a$. Assuming that ion $b$ at point $\mathbf{r}_b$ causes a rather small perturbation of the solution environment, we obtain in the linear approximation in concentration perturbations
\begin{equation}
\label{chem_pot_lin}
\mu_{a}(\mathbf{r}_a,\mathbf{r}_b)\approx \bar{\mu}_{a}+k_B Th_{ab}(\mathbf{r}_a,\mathbf{r}_b)-k_{B}T \sum_{l}\int d^3\mathbf{r}'\,c_{al}(\mathbf{r}_a-\mathbf{r}')\,\delta c_{l}(\mathbf{r}',\mathbf{r}_b),
\end{equation} 
where $c_{al}$ is the bulk direct correlation function and $\bar{\mu}_a$ is the bulk chemical potential, $h_{ab}(\mathbf{r}_a,\mathbf{r}_b)=g_{ab}(\mathbf{r}_a,\mathbf{r}_b)-1$ is the total correlation function, and the concentration perturbations are given by
\begin{equation}
\label{delta_c}
\delta c_{a}(\mathbf{r},\mathbf{r}')=
\begin{cases}
\bar{c}_{a}h_{ab}(\mathbf{r},\mathbf{r}'), & a\neq b,\\[2pt]
\bar{c}_{b}h_{bb}(\mathbf{r},\mathbf{r}')+\delta(\mathbf{r}-\mathbf{r}'), & a= b.
\end{cases}
\end{equation}

Writing the non-equilibrium total correlation functions as a sum of equilibrium and steady-state parts,
\begin{equation}
\label{h_tot}
h_{ab}(\mathbf{r},\mathbf{r}') = \bar{h}_{ab}(\mathbf{r} - \mathbf{r}') + \omega_{ab}(\mathbf{r},\mathbf{r}'),
\end{equation}
we note that the equilibrium functions $\bar{h}_{ab}$ are independent of the magnetic field. This is guaranteed by the Bohr--van Leeuwen theorem \cite{kaufman2025lectures} of classical statistical mechanics and is consistent with molecular dynamics simulations of electrolyte solutions in magnetic fields.
Now, substituting (\ref{h_tot}) into the continuity equation \eqref{continuous}, and then using the Ornstein-Zernike relation
\begin{equation}
\label{OZ}
\bar{h}_{ab}(\mathbf{r}_a-\mathbf{r}_b)= c_{ab}(\mathbf{r}_a-\mathbf{r}_b)+\int d^3\mathbf{r}'\sum_{l}c_{al}(\mathbf{r}_a-\mathbf{r}') \bar{c}_{l} \bar{h}_{lb}(\mathbf{r}'-\mathbf{r}_b),
\end{equation}
we arrive at (for more details, see Appendix \ref{app:derivation_basic})
\begin{multline}
\label{basic}
m_a k_B T \nabla_a^2 \hat{\omega}_{ab}(\mathbf{r}_a,\mathbf{r}_b)+m_b k_B T \nabla_b^2 \hat{\omega}_{ba}(\mathbf{r}_b,\mathbf{r}_a)\\
-\beta_a m_a  q_a \mathbf{B}\cdot[\nabla_a\times \mathbf{u}_b(\mathbf{r}_a,\mathbf{r}_b)]
-\beta_b m_b q_b \mathbf{B}\cdot[\nabla_b\times \mathbf{u}_a(\mathbf{r}_b,\mathbf{r}_a)]\\
=(m_a\mathbf{f}_{a}(\mathbf{r}_a,\mathbf{r}_b)+\mathbf{u}_b(\mathbf{r}_a,\mathbf{r}_b))\cdot\nabla_a h_{ab}(\mathbf{r}_a,\mathbf{r}_b)\\
+(m_b\mathbf{f}_{b}(\mathbf{r}_b,\mathbf{r}_a)+\mathbf{u}_a(\mathbf{r}_b,\mathbf{r}_a))\cdot\nabla_b h_{ba}(\mathbf{r}_b,\mathbf{r}_a).
\end{multline}
Here we have used that in the linear approximation with respect to $\mathbf{B}$
\begin{align}
\nabla_a\cdot \mathbf{f}_a(\mathbf{r}_a,\mathbf{r}_b)&\simeq \beta_a q_a \mathbf{B}\cdot[\nabla_a\times \mathbf{u}_b(\mathbf{r}_a,\mathbf{r}_b)],\\
\nabla_b\cdot \mathbf{f}_b(\mathbf{r}_b,\mathbf{r}_a)&\simeq \beta_b q_b \mathbf{B}\cdot[\nabla_b\times \mathbf{u}_a(\mathbf{r}_b,\mathbf{r}_a)],
\end{align}
introduced the shorthand notation
\begin{equation}
\hat{\omega}_{ab}(\mathbf{r}_a,\mathbf{r}_b)=\omega_{ab}(\mathbf{r}_a,\mathbf{r}_b)-\int d^3\mathbf{r}'\sum_{l}c_{al}(\mathbf{r}_a-\mathbf{r}') \bar{c}_{l} \omega_{lb}(\mathbf{r}',\mathbf{r}_b),
\end{equation}
and taken into account that in the linear approximation in $h_{ab}$ in \eqref{continuous} $g_{ab}(\mathbf{r}_a,\mathbf{r}_b)\approx 1$, together with the incompressibility condition
\begin{equation}
\label{incompres}
\nabla_a \cdot\mathbf{u}_b(\mathbf{r}_a,\mathbf{r}_b)=\nabla_b \cdot\mathbf{u}_a(\mathbf{r}_b,\mathbf{r}_a)=0.
\end{equation}

\subsubsection{Hydrodynamic response.}

To close the system \eqref{basic}, we need a hydrodynamic equation for the velocity field. For steady state and relatively weak solvent flow, the linearized Navier-Stokes equation is appropriate:
\begin{equation}
\label{Stokes}
\eta \nabla_a^2 \mathbf{u}_b(\mathbf{r}_a,\mathbf{r}_b)-\nabla_a p_b(\mathbf{r}_a,\mathbf{r}_b)+\mathbf{E} \varrho_b(\mathbf{r}_a,\mathbf{r}_b)+[\mathbf{J}_b(\mathbf{r}_a,\mathbf{r}_b)\times \mathbf{B}]=0,\quad a \leftrightarrow b,
\end{equation}
where $p_b(\mathbf{r}_a, \mathbf{r}_b)$ is the solvent pressure at $\mathbf{r}_a$ near ion $b$. The magnetic force density is taken proportional to $\alpha$, which is the dielectric friction coefficient of the solvent; this choice is made for simplicity and because the electrophoretic contribution is subdominant compared with the correlation and mean-field parts. The local ionic charge density is
\begin{equation}
\varrho_b(\mathbf{r}_a,\mathbf{r}_b)=\sum_{l}\bar{c}_l\hat{q}_l h_{lb}(\mathbf{r}_a,\mathbf{r}_b),\quad a \leftrightarrow b,
\end{equation}
and we have introduced the charge smearing integral operator~\cite{budkov2026conductivity}
\begin{equation}
\hat{q}_l h_{lb}(\mathbf{r}_a,\mathbf{r}_b)=\int d^3\mathbf{r}' \hat{q}_{l}(\mathbf{r}_a-\mathbf{r}')h_{lb}(\mathbf{r}',\mathbf{r}_b),
\end{equation}
whose kernel $\hat{q}_{a}(\mathbf{r})$ represents the charge density of ion $a$, normalized as $\int d^3\mathbf{r}\,\hat{q}_{a}(\mathbf{r}) =q_a$. The corresponding current density is
\begin{equation}
\mathbf{J}_b(\mathbf{r}_a,\mathbf{r}_b)=\sum_{l}\beta_l\bar{c}_l\hat{q}_l h_{lb}(\mathbf{r}_a,\mathbf{r}_b)\mathbf{v}_{lb}(\mathbf{r}_a,\mathbf{r}_b).
\end{equation}

For a homogeneous bulk solution it is useful to decompose the solvent response into electric and magnetic parts. Exploiting translational invariance and approximating the ion velocity field in the magnetic force density as $\mathbf{v}_{ab} \simeq m_a q_a \mathbf{E}$, which retains the leading equilibrium-cloud contribution at the pair-correlation accuracy used here, one obtains in Fourier space
\begin{equation}
\label{NS_el}
-\eta k^2 \mathbf{u}_b^{\text{E}}(\mathbf{k})-i\mathbf{k}p_b^{\text{E}}(\mathbf{k})+\mathbf{E}\bar{\varrho}_b(\mathbf{k})=0,
\end{equation}
\begin{equation}
\label{NS_magn}
-\eta k^2 \mathbf{u}_b^{\text{H}}(\mathbf{k})-i\mathbf{k}p_b^{\text{H}}(\mathbf{k})+[\bar{\mathbf{J}}_{b}(\mathbf{k})\times \mathbf{B}]=0,
\quad
\bar{\mathbf{J}}_{b}(\mathbf{k})\simeq \sum_{a}\beta_a m_a\bar{c}_a q_a\hat{q}_a(\mathbf{k})\bar{h}_{ab}(\mathbf{k}) \mathbf{E}.
\end{equation}
Projecting Eqs.~\eqref{NS_el} and \eqref{NS_magn} onto the subspace transverse to $\mathbf{k}$, as in the standard Oseen solution of the incompressible Stokes equation~\cite{kim2005microhydrodynamics,pitaevskii2012physical}, gives
\begin{equation}
\label{velocity_el}
\mathbf{u}_b^{\text{E}}(\mathbf{k})=\mathbf{u}_b^{\text{E}}(-\mathbf{k})=
\frac{1}{\eta k^2}
\left(\mathbf{E}-\frac{\mathbf{k}(\mathbf{k}\cdot \mathbf{E})}{k^2}\right)\bar{\varrho}_b(\mathbf{k}),
\end{equation}
\begin{equation}
\label{velocity_hall}
\mathbf{u}_b^{\text{H}}(\mathbf{k})=
\frac{1}{\eta k^2}
\left(\mathbf{1}-\frac{\mathbf{k}\mathbf{k}}{k^2}\right)\cdot
[\bar{\mathbf{J}}_b(\mathbf{k})\times\mathbf{B}].
\end{equation}
Thus the electric field produces the ordinary electrophoretic back-flow, whereas the magnetic field rotates the ionic current density and generates a vortical Hall component of the solvent velocity. The solution used here is the bulk, unbounded Oseen solution of the incompressible Stokes problem: the disturbance velocity is required to vanish at infinity and no cell-wall boundary condition is imposed. Consequently, the theory gives the intrinsic bulk Hall conductivity. Possible geometry-dependent hydrodynamic corrections associated with finite experimental cells or nearby walls are outside the scope of the present bulk formulation.

\subsubsection{Linearized Hall distortion of the ionic atmosphere.}

In the linear approximation with respect to $h_{ab}$ on the right-hand side of \eqref{basic}, we may neglect the solvent velocity field $\mathbf{u}_b$, because it is itself linear in $h_{ab}$ and would generate quadratic terms. Physically, this corresponds to ignoring the influence of electrophoretic flow on the steady-state correlation functions, as was done in the classical DHO theory. In this approximation the correlation functions become translationally invariant, $\omega_{ab}(\mathbf{r}_a,\mathbf{r}_b)=\omega_{ab}(\mathbf{r}_a-\mathbf{r}_b)$, and we may replace $h_{ab}$ by its equilibrium part, $h_{ab}(\mathbf{r}_a,\mathbf{r}_b)=\bar{h}_{ab}(\mathbf{r}_a-\mathbf{r}_b)$. Furthermore, we split the steady-state correlation function into electric (E) and magnetic, or Hall (H), contributions,
\begin{equation}
\omega_{ab}(\mathbf{r})=\omega_{ab}^{\text{E}}(\mathbf{r})+\omega_{ab}^{\text{H}}(\mathbf{r}),
\end{equation}
where $\omega_{ab}^{\text{E}}(\mathbf{r})\propto \mathbf{E}$ is described by the existing SCDHO theory of electric conductivity~\cite{budkov2026conductivity}, while $\omega_{ab}^{\text{H}}(\mathbf{r})\propto \mathbf{E}\times\mathbf{B}$ satisfies the following equation in Fourier representation:
\begin{multline}
\label{corr_funct_hall}
m_a k_B T k^2 \hat{\omega}_{ab}^{\text{H}}(\mathbf{k})+m_b k_B T k^2 \hat{\omega}_{ba}^{\text{H}}(-\mathbf{k})\\
=-\Big[(\beta_a m_a^2 q_a^2-\beta_b m_b^2 q_b^2)\bar{h}_{ab}(\mathbf{k})+\frac{1}{\eta k^2}\Bigl(\beta_a q_a m_a \bar{\varrho}_b(\mathbf{k})-\beta_b q_b m_b \bar{\varrho}_a(\mathbf{k})\Bigr)\Big] i\mathbf{k}\!\cdot\![\mathbf{E}\times\mathbf{B}],
\end{multline}
with the shorthand
\begin{equation}
\hat{\omega}_{ab}^{\text{H}}(\mathbf{k})={\omega}_{ab}^{\text{H}}(\mathbf{k})-\sum_{l}c_{al}(\mathbf{k})\bar{c}_{l}\omega_{lb}^{\text{H}}(\mathbf{k})
\end{equation}
and the equilibrium Fourier transform of the local charge density around ion $a$,
\begin{equation}
\bar{\varrho}_a(\mathbf{k})=\sum_{l}\bar{c}_l\hat{q}_l(\mathbf{k}) \bar{h}_{la}(\mathbf{k}).
\end{equation}
In deriving \eqref{corr_funct_hall} we have used the electric solvent velocity \eqref{velocity_el}. Because $\omega_{ab}^{\text{H}}(\mathbf{k})$ are odd functions with the structure $\omega_{ab}^{\text{H}}(\mathbf{k})=\mathbf{k}\cdot[\mathbf{E}\times\mathbf{B}]f_{ab}(k)$ and must satisfy $\omega_{ab}^{\text{H}}(\mathbf{k})=\omega_{ba}^{\text{H}}(-\mathbf{k})$, we obtain the antisymmetry condition
\begin{equation}
\label{antisymmetry}
{\omega}_{ab}^{\text{H}}(\mathbf{k})=-{\omega}_{ba}^{\text{H}}(\mathbf{k}),
\end{equation}
which implies $\omega_{aa}^{\text{H}}(\mathbf{k}) = 0$.

\subsubsection{Hall current and transport coefficients.}

The Lorentz force in Eq.~\eqref{velocity} acts on the full ionic velocity. To first order in $\mathbf B$, write the field-induced chemical potential and solvent velocity as $\delta\mu_a=\delta\mu_a^E+\delta\mu_a^H$ and $\mathbf u_b=\mathbf u_b^E+\mathbf u_b^H$, where $E$ and $H$ denote terms proportional to $E$ and $EB$, respectively. The corresponding conditional ionic velocities are
\begin{align}
\mathbf v_{ab}^{E}
&=m_aq_a\mathbf E-m_a\nabla_a\delta\mu_a^E+\mathbf u_b^E,
\label{velocity_full_E}\\
\mathbf v_{ab}^{H}
&=\beta_am_a^2q_a^2[\mathbf E\times\mathbf B]
-m_a\nabla_a\delta\mu_a^H+\mathbf u_b^H\notag\\
&\quad-\beta_am_a^2q_a[\nabla_a\delta\mu_a^E\times\mathbf B]
+\beta_am_aq_a[\mathbf u_b^E\times\mathbf B].
\label{velocity_full_H}
\end{align}
The last two terms are the Lorentz rotations of the relaxation velocity and the ordinary electrophoretic back-flow. Both must be retained when constructing the Hall current, even though the linearized equation for the Hall distortion of the ionic atmosphere is unchanged at the present accuracy.

The total average current density is
\begin{equation}
\mathbf J=\sum_aq_a\bar c_a\mathbf v_a
=\mathbf J^E+\mathbf J^H,
\qquad \mathbf J^E=\sigma\mathbf E,
\end{equation}
where $\mathbf v_a=\lim_{\mathbf r'\to\mathbf r}\mathbf v_{aa}(\mathbf r-\mathbf r')$ and $\sigma$ is the ordinary conductivity obtained in Ref.~\cite{budkov2026conductivity}. The complete Hall current has five contributions,
\begin{equation}
\mathbf J^H=\mathbf J_{\rm MF}^H+\mathbf J_{\rm atm}^H
+\mathbf J_{\rm rot,cor}^H+\mathbf J_{\rm vort}^H
+\mathbf J_{\rm rot,ep}^H.
\label{hall_five_terms}
\end{equation}
The mean-field and atmosphere-distortion currents are
\begin{align}
\mathbf J_{\rm MF}^H
&=\sum_a\beta_am_a^2q_a^3\bar c_a[\mathbf E\times\mathbf B],\\
\mathbf J_{\rm atm}^H
&=k_BT\intk i\mathbf k
\sum_{a,b\ne a}q_am_a\bar c_a\bar c_b
c_{ab}(\mathbf k)\omega_{ab}^H(\mathbf k).
\end{align}
The species-resolved ordinary relaxation current is
\begin{equation}
\mathbf J_{{\rm cor},a}^E
=k_BTq_am_a\bar c_a\intk i\mathbf k
\sum_{b\ne a}\bar c_b c_{ab}(\mathbf k)\omega_{ab}^E(\mathbf k).
\end{equation}
Its Lorentz rotation produces
\begin{equation}
\mathbf J_{\rm rot,cor}^H
=\sum_a\beta_am_aq_a
[\mathbf J_{{\rm cor},a}^E\times\mathbf B].
\end{equation}
The two electrophoretic Hall currents are
\begin{align}
\mathbf J_{\rm vort}^H
&=\frac{2[\mathbf E\times\mathbf B]}{3\eta}
\intk\frac{1}{k^2}\sum_{a,b}
\beta_am_aq_a\hat q_a(k)q_b\bar c_a\bar c_b\bar h_{ab}(k),
\label{hall_vort_current}\\
\mathbf J_{\rm rot,ep}^H
&=\frac{2[\mathbf E\times\mathbf B]}{3\eta}
\intk\frac{1}{k^2}\sum_{a,b}
\beta_am_aq_a^2\hat q_b(k)\bar c_a\bar c_b\bar h_{ab}(k).
\label{hall_rot_ep_current}
\end{align}
The first is generated by the vortical magnetic solvent response; the second is the Lorentz rotation of the ordinary species-resolved electrophoretic current,
\begin{equation}
\mathbf J_{{\rm ep},a}^E
=\frac{2q_a\bar c_a\mathbf E}{3\eta}
\intk\frac{\sum_b\bar c_b\bar h_{ab}(k)\hat q_b(k)}{k^2}.
\end{equation}
We group the interaction terms as
\begin{equation}
\mathbf J_{\rm cor}^H=\mathbf J_{\rm atm}^H+\mathbf J_{\rm rot,cor}^H,
\qquad
\mathbf J_{\rm ph}^H=\mathbf J_{\rm vort}^H+\mathbf J_{\rm rot,ep}^H,
\end{equation}
so that the usual three-channel decomposition remains valid when each channel includes both of its vertices.

The Hall conductivity $\sigma^H$ is defined by
\begin{equation}
\label{sigma}
\mathbf J^H=\sigma^H[\mathbf E\times\mathbf B].
\end{equation}
Thus the coefficient $\sigma^H$ does not include the magnetic field itself. The conductivity tensor, determined by $J_\alpha=\sigma_{\alpha\beta}E_\beta$, is~\cite{landau2013electrodynamics}
\begin{equation}
\sigma_{\alpha\beta}=\sigma\delta_{\alpha\beta}
+\sigma^H\varepsilon_{\alpha\beta\gamma}B_\gamma,
\end{equation}
where repeated Cartesian indices are summed. In the weak-field limit the Hall constant is~\cite{sung1987microscopic}
\begin{equation}
R_{\rm H}=\frac{\sigma^H}{\sigma^2}.
\end{equation}
For a 1:1 electrolyte, the dimensionless Hall number is
\begin{equation}
\label{hall_number_def}
h=R_{\rm H}nq,\qquad n=\bar c_1=\bar c_2,\qquad q=e.
\end{equation}
Here $n$ is the number density of electrolyte formula units. Concentrations in the theoretical equations are number densities; the molarity $c$ used on the numerical plots is related to them by $n=10^3N_Ac$ when $c$ is expressed in mol\,l$^{-1}$ and $n$ in m$^{-3}$, with $N_A$ the Avogadro constant.

Finally, we note that throughout this work we neglect possible quantum-mechanical effects of the magnetic field on the internal charge structure of the ions. In the classical regime relevant to typical laboratory conditions ($B\lesssim 1$~T), the Slater-type form factors $\hat{q}_a(\mathbf{k})$ can be considered independent of $\mathbf{B}$, as any field-induced deformation of the ionic charge clouds is negligible compared to thermal fluctuations.

\subsection{Hall effect in one-component electrolyte solutions}

\subsubsection{General theory.}
For a binary electrolyte, let $q_1=z_1e$ and $q_2=-z_2e$, with $z_1,z_2>0$ and $z_1\bar c_1=z_2\bar c_2$. The only nonzero Hall correlation function is
\begin{equation}
\omega_{12}^H(\mathbf k)=-\omega_{21}^H(\mathbf k),
\qquad \omega_{11}^H=\omega_{22}^H=0.
\end{equation}
Introduce
\begin{align}
d&=m_1q_1-m_2q_2,\qquad
A=\beta_1m_1^2q_1^2-\beta_2m_2^2q_2^2,\\
D(k)&=m_1[1-\bar c_1c_{11}(k)]+m_2[1-\bar c_2c_{22}(k)],\\
\Delta(k)&=1-\bar c_1c_{11}(k)-\bar c_2c_{22}(k)
+\bar c_1\bar c_2[c_{11}(k)c_{22}(k)-c_{12}^2(k)],\\
H_{\rm hyd}(k)&=\frac{\beta_1q_1m_1\bar\varrho_2(k)-\beta_2q_2m_2\bar\varrho_1(k)}{\eta k^2}.
\end{align}
The equilibrium Ornstein--Zernike relation gives $\bar h_{12}=c_{12}/\Delta$. The ordinary and Hall distortions are
\begin{align}
\omega_{12}^E(\mathbf k)
&=-\frac{d\,i(\mathbf E\cdot\mathbf k)c_{12}(k)}
{k_BT k^2\Delta(k)D(k)},
\label{binary_omega_E}\\
\omega_{12}^H(\mathbf k)
&=-\frac{i\mathbf k\cdot[\mathbf E\times\mathbf B]}{k_BT k^2D(k)}
\left[A\bar h_{12}(k)+H_{\rm hyd}(k)\right].
\label{binary_omega_H}
\end{align}
Equation~\eqref{binary_omega_H} follows from Eq.~\eqref{corr_funct_hall}. Its hydrodynamic term is produced by the coupling of the magnetic field to the vorticity of the electric-field-induced solvent flow.

The ordinary relaxation correction and its Lorentz rotation are
\begin{align}
\Delta\sigma_{\rm cor}^E
&=-\frac{d^2\bar c_1\bar c_2}{3}
\intk\frac{c_{12}^2(k)}{\Delta(k)D(k)},\\
\Delta\sigma_{\rm rot,cor}^H
&=-\frac{Ad\bar c_1\bar c_2}{3}
\intk\frac{c_{12}^2(k)}{\Delta(k)D(k)}.
\end{align}
The nonhydrodynamic part of the atmosphere current equals this rotated-relaxation contribution. The complete correlation correction therefore reads
\begin{equation}
\Delta\sigma_{\rm cor}^H
=-\frac{d\bar c_1\bar c_2}{3}
\intk\frac{c_{12}(k)}{D(k)}
\left[2A\bar h_{12}(k)+H_{\rm hyd}(k)\right].
\label{binary_sigma_cor}
\end{equation}
Only the nonhydrodynamic relaxation term is doubled; the hydrodynamic mobility-mismatch term is unchanged.

Combining Eqs.~\eqref{hall_vort_current} and \eqref{hall_rot_ep_current} gives the complete electrophoretic correction,
\begin{equation}
\Delta\sigma_{\rm ph}^H
=\frac{2}{3\eta}\intk\frac{1}{k^2}
\sum_{a,b=1}^2\beta_am_aq_a\bar c_a\bar c_b\bar h_{ab}(k)
\left[q_b\hat q_a(k)+q_a\hat q_b(k)\right].
\label{binary_sigma_ep}
\end{equation}
These two terms coincide for point charges, but need not coincide for different finite charge distributions.

The total Hall conductivity is
\begin{equation}
\sigma^H=\sigma_{\rm MF}^H+\Delta\sigma_{\rm cor}^H+\Delta\sigma_{\rm ph}^H,
\end{equation}
where
\begin{equation}
\sigma_{\rm MF}^H=e^3
\left(\beta_1m_1^2z_1^3\bar c_1-\beta_2m_2^2z_2^3\bar c_2\right).
\end{equation}

\subsubsection{Model of smeared charges.} 
To proceed further we need the direct correlation functions and the charge form factors. We represent the charge density of ion $a$ as $\hat q_a(\mathbf r)=q_a\rho_a(r)$, where the normalized Slater-type distribution is~\cite{warren2014screening,chiacchiera2024coarse}
\begin{equation}
\rho_a(r)=\frac{1}{8\pi l_a^3}\exp\left(-\frac{r}{l_a}\right),
\qquad
\int d^3\mathbf r\,\rho_a(r)=1 .
\end{equation}
Its Fourier transform is
\begin{equation}
\hat q_a(k)=q_a\rho_a(k),
\qquad
\rho_a(k)=\frac{1}{(1+k^2l_a^2)^2}.
\end{equation}
Here $l_a$ is the effective charge-smearing length. It characterizes the spatial range over which the ionic charge is distributed in the coarse-grained electrostatic model; it should not be confused with a bare crystallographic radius or with a hydrodynamic Stokes radius. The smearing regularizes only the electrostatic part of the interaction and does not replace the short-range repulsion associated with excluded volume, exchange effects, or specific solvation. These non-electrostatic effects are neglected at the RPA level and should be incorporated through more refined closures when strongly overlapping solvation shells become important.

In the random phase approximation (RPA) with smeared charges,
\begin{equation}
c_{ab}(\mathbf{k})=-\frac{\hat{q}_a(\mathbf{k})\hat{q}_b(\mathbf{k})}{k_B T\varepsilon k^2},
\end{equation}
so that
\begin{equation}
\bar{h}_{ab}(\mathbf{k})=-\frac{\hat{q}_a(\mathbf{k})\hat{q}_b(\mathbf{k})}{k_B T\varepsilon \bigl(k^2+\varkappa^2(k)\bigr)},
\end{equation}
where the screening function is
\begin{equation}
\varkappa^2(k)=\frac{1}{\varepsilon k_{B}T}\sum_{a}\bar{c}_a\hat{q}_a(k)^2,
\end{equation}
and consequently
\begin{equation}
\bar{\varrho}_{a}(\mathbf{k})=-\hat{q}_a(\mathbf{k})\frac{\varkappa^2(k)}{k^2+\varkappa^2(k)}.
\end{equation}
With $\hat q_1(k)=z_1e\rho_1(k)$ and $\hat q_2(k)=-z_2e\rho_2(k)$, the correlation contribution becomes
\begin{equation}
\label{delta_sigma_cor}
\Delta \sigma_{\text{cor}}^{\text{H}}=-\frac{(z_1z_2)^2 e^7\bar{c}_1\bar{c}_2(z_1m_1+z_2 m_2)}{6\pi^2(\varepsilon k_B T)^2(m_1+m_2)}\int_{0}^{\infty}\frac{dk\,k^2\rho_1(k)\rho_2(k)\,W(k)}{(k^2+\varkappa^2(k))(k^2+\varkappa_1^2(k))},
\end{equation}
where
\begin{equation}
\varkappa^2(k)=\frac{e^2}{\varepsilon k_B T}\bigl(z_1^2\rho_1^2(k)\bar{c}_1+z_2^2\rho_2^2(k)\bar{c}_2\bigr),\qquad
\varkappa_1^2(k)=\frac{e^2}{\varepsilon k_B T}\frac{z_1^2\rho_1^2(k)\bar{c}_1m_1+z_2^2\rho_2^2(k)\bar{c}_2m_2}{m_1+m_2},
\end{equation}
and the auxiliary function
\begin{multline}
W(k)=2(\beta_1 m_1^2 z_1^2-\beta_2 m_2^2 z_2^2)\rho_1(k)\rho_2(k)\\
+\frac{\bigl(z_1^2\rho_1^2(k)\bar{c}_1+z_2^2\rho_2^2(k)\bar{c}_2\bigr)\bigl(\beta_1 m_1\rho_2(k)-\beta_2 m_2\rho_1(k)\bigr)}{\eta k^2}.
\end{multline}

The electrophoretic contribution to the Hall conductivity is
\begin{equation}
\label{delta_sigma_ph}
\Delta\sigma_{\rm ph}^H
=-\frac{e^5}{3\pi^2\eta\varepsilon k_BT}
\int_0^\infty dk\,
\frac{I_1(k)\Lambda_2(k)+I_2(k)\Lambda_1(k)}{k^2+\varkappa^2(k)},
\end{equation}
where
\begin{align}
I_1(k)&=z_1^2\bar c_1\rho_1(k)+z_2^2\bar c_2\rho_2(k),\\
I_2(k)&=z_1^2\bar c_1\rho_1^2(k)+z_2^2\bar c_2\rho_2^2(k),\\
\Lambda_1(k)&=\beta_1m_1\bar c_1z_1^3\rho_1(k)
-\beta_2m_2\bar c_2z_2^3\rho_2(k),\\
\Lambda_2(k)&=\beta_1m_1\bar c_1z_1^3\rho_1^2(k)
-\beta_2m_2\bar c_2z_2^3\rho_2^2(k).
\end{align}
The products $I_1\Lambda_2$ and $I_2\Lambda_1$ arise from the vortical and rotated-electrophoretic currents, respectively. For unequal finite form factors, the second term cannot in general be included by simply doubling the first.

\subsubsection{Limit of point-like charges.} 
In the point-charge limit ($\rho_a(k)=1$), appropriate for the dilute-limit benchmark, integrating over $k$ in~\eqref{delta_sigma_cor} and~\eqref{delta_sigma_ph} yields the following analytic expressions:
\begin{equation}
\label{delta_sigma_cor_point}
\Delta \sigma_{\text{cor}}^{\text{H}}=-\frac{e^7\bar{c}_1\bar{c}_2(z_1z_2)^2(z_1m_1+z_2 m_2)}{12\pi (\varepsilon k_B T)^2(m_1+m_2)(\kappa+\kappa_1)}\Bigl(2(\beta_1 m_1^2 z_1^2-\beta_2 m_2^2 z_2^2) + \frac{2I (\beta_1 m_1-\beta_2 m_2)}{\kappa\kappa_1 \eta}\Bigr),
\end{equation}
\begin{equation}
\label{delta_sigma_ph_point}
\Delta \sigma_{\text{ph}}^{\text{H}}=-\frac{2e^5 I (\beta_1 m_1 z_1^3 \bar{c}_1 - \beta_2 m_2 z_2^3\bar{c}_2)}{3\pi\varepsilon k_B T \eta\kappa}.
\end{equation}
Here $\kappa=\bigl(2 e^2I/\varepsilon k_B T\bigr)^{1/2}$ is the inverse Debye length with the standard ionic strength, $I=(z_1^2\bar{c}_1+z_2^2\bar{c}_2)/2$, and $\kappa_1=\bigl(2e^2I_{\text{m}}/\varepsilon k_B T\bigr)^{1/2}$ is the steady-state inverse screening length with the mobility-weighted ionic strength, $I_{\text{m}}=(z_1^2m_1\bar{c}_1+z_2^2m_2\bar{c}_2)/[2(m_1+m_2)]$.

The second term inside the brackets of \eqref{delta_sigma_cor_point} originates from the $\bar{\varrho}$-dependent part of the Hall correlation function~\eqref{corr_funct_hall} -- the `ion mobility mismatch' contribution -- and has no analogue in the Ohmic conductivity~\cite{pitaevskii2012physical}. In the standard DHO theory~\cite{fuoss1978review}, the influence of the solvent flow on the ionic correlation functions is absent in the linear approximation, and the electrophoretic effect enters only through the average velocity field~\cite{pitaevskii2012physical}. By contrast, the Hall geometry couples the vortical solvent motion induced by the magnetic field to the interionic correlations already at linear order in the correlation functions. This cross-coupling is therefore a distinctive and physically unavoidable feature of magnetotransport in electrolytes.

Setting $\rho_a(k)=1$ removes the short-range electrostatic regularization. For fixed stoichiometry, the mean-field Hall conductivity is proportional to the number density $n$, whereas both interaction corrections scale as $n^{3/2}$. This is the classical limiting-law structure. Equality of the full hydrodynamic result with every term of Friedman's cluster expansion is not asserted here.

A direct analytical check is available when hydrodynamics is suppressed. Set $\eta^{-1}=0$ and $\beta_+=\beta_-=1$, and consider a symmetric $z:z$ electrolyte, $q_+=q=-q_-$ and $\bar c_+=\bar c_-=n$. The off-diagonal conductivity, now containing $B$ explicitly, obeys
\begin{equation}
\Delta\kappa_H
=2qB(m_+-m_-)\Delta\kappa_{\rm DHO}+O(B^3),
\label{avni_limit_check}
\end{equation}
where $\Delta\kappa_{\rm DHO}$ is the ordinary DHO relaxation correction. With $m_\pm=\gamma_\pm^{-1}$, this is Eq.~(86b) of the independent odd stochastic density functional theory calculation of Avni \textit{et al.}~\cite{avni2026hall}. It also agrees with the relaxation part of Friedman's result~\cite{friedman1965calculation}. This comparison tests the nonhydrodynamic relaxation sector; the hydrodynamic terms in Eqs.~\eqref{delta_sigma_cor_point} and \eqref{delta_sigma_ph_point} are additional predictions within the present solvent approximation.

\section{Numerical results}

\subsection{Test of "infinite dilution"}
To assess the overall performance of the SCDHO theory with ion-specific dielectric friction coefficients $\beta_a$, we compare the calculated Hall constant, $R_H$, and the dimensionless Hall number, $h$, with the experimental data reported by Meton and Gerard for a set of dilute aqueous solutions of 1:1 electrolytes \cite{meton1976hall}. The Stokes radii were taken from ref. \cite{nightingale1959phenomenological}, the electrical conductivities at infinite dilution -- from ref. \cite{Holze2016Electrochemistry}. As the reported data correspond to concentrations of several mM, the charge-smearing parameters are not required and the point-charge limit of the theory is sufficient. The experimental uncertainty is point-specific and has been propagated from the stated measurement noise of $\Delta V_H\approx20~nV$.

\begin{figure}
\centering
\includegraphics[width=\linewidth]{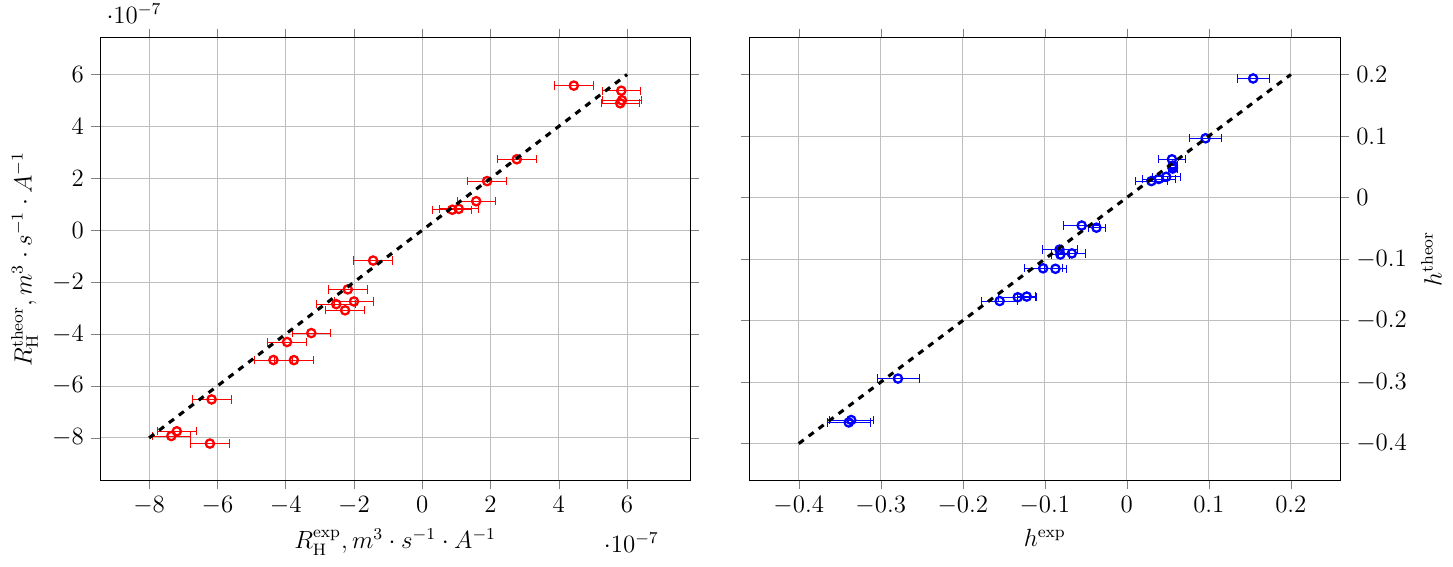}
\caption{Parity plots for the Hall constant $R_{\rm H}$ (left) and dimensionless Hall number $h$ (right) for 22 dilute aqueous 1:1 electrolytes. Theory is evaluated in the point-charge limit. Experimental data are from Ref.~\cite{meton1976hall}; error bars are propagated from its reported Hall-voltage uncertainty $\Delta V_{\rm H}\simeq20$ nV.}
\label{parity_plots}
\end{figure}

Figure~\ref{parity_plots} compares the Hall constants and Hall numbers for 22 dilute electrolytes. With all charge-smearing lengths set to zero, the mean signed relative deviations,
\begin{equation}
\frac{1}{N}\sum_i\frac{y_i^{\rm theor}-y_i^{\rm exp}}{|y_i^{\rm exp}|},
\end{equation}
are $-12.8\%$ for $R_{\rm H}$ and $-11.8\%$ for $h$. The corresponding mean absolute relative deviations are $16.8\%$ and $16.9\%$. Twelve of the 22 systems fall within the experimental uncertainties propagated from the Hall-voltage accuracy. Since the data include both signs, a negative signed deviation does not by itself imply that every signal magnitude is underestimated. The theory captures the signs and typical magnitudes, while quantitative discrepancies remain.

\subsection{\texorpdfstring{Case 1: $\beta_+\approx\beta_-$, $m_+\approx m_-$ -- Salt solutions}{Case 1: Salt solutions}}

We first apply the model to two prototypical aqueous salt solutions, NaI and NaBPh$_4$, for which we can find the experimental concentration dependence of the Hall number \cite{abbes1980effet}. In both systems the cation and anion have comparable Stokes mobilities ($m_{\text{I}^-}/m_{\text{Na}^+}\approx1.5$ and $m_{\text{Na}^+}/m_{\text{BPh}_4^-}\approx2.5$) and ion-specific dielectric friction coefficients $\beta_a$ (ionic Hall numbers) are of similar magnitude (see Table \ref{input_parameters}). The effective charge-smearing lengths were fitted to the Ohmic conductivity data (Table \ref{input_parameters}).

\begin{table}
    \centering
    \begin{tabular}{l|c|c|l|c|c}
         Ion     & $l_a^{\rm eff}$, [\angstrom] & $\beta_a, [-]$ & Ion       & $l_a^{\rm eff}$, [\angstrom] & $\beta_a, [-]$\\
         \hline
         K$^+$   & 0.6                &   0.8     &  I$^-$    &   5.5              &  0.6     \\
         Li$^+$  & 0.3                &   1       &  NO$_3^-$ &   1                &  0.6     \\              
         Na$^+$  & 0.4                &   0.9     &  Br$^-$   &   0.5              &  0.6     \\
         Hmim$^+$& 0.1                &   0.7     &  Cl$^-$   &   0.8              &  0.5     \\
         H$^+$   & 0.5                &   0.1     &  BPh$_4^-$&   2                &  1       \\
                 &                    &           &  OH$^-$   &   0.8              &  0.2     \\       
    \end{tabular}
    \caption{Ions, effective charge-smearing lengths $l_a$, and ionic Hall numbers $\beta_a$ used in the calculations. The parameter $l_a$ specifies the width of the electrostatic form factor and should not be interpreted as a bare or solvated ionic radius. In particular, the small value used for Hmim$^+$ reflects the effective localization of the coarse-grained charge distribution, not the steric size of the imidazolium cation.}
    \label{input_parameters}
\end{table}

\begin{figure}
    \centering
    \includegraphics[width=0.6\linewidth]{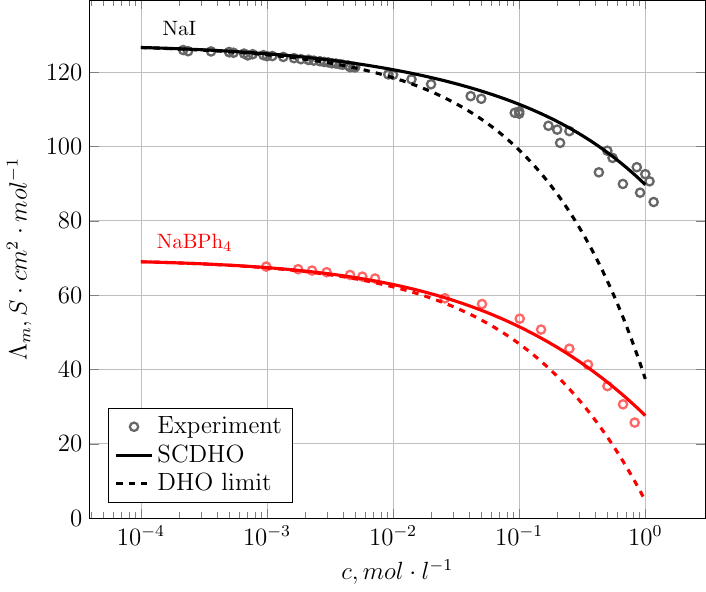}
    \caption{Concentration dependence of molar electric conductivity of NaI and NaBPh$_4$ solutions. Solid lines: smeared charges (SCDHO); dashed lines: classical DHO limit. Experimental data from refs.~\cite{Holze2016Electrochemistry, schiavo1977conductimetric,wachter2006hydration}.}
    \label{electric_conductivity_salts}
\end{figure}

Figure \ref{electric_conductivity_salts} displays the molar electric conductivity $\Lambda_m$ as a function of concentration for two salts. The experimental data (symbols) are taken from refs. \cite{Holze2016Electrochemistry, schiavo1977conductimetric,wachter2006hydration} and cover the range from $10^{-4}$ to approximately 1 M. The solid lines correspond to the SCDHO predictions with charge smearing, while the dashed lines represent the classical Debye-H\"uckel-Onsager (DHO) point-charge limit ($l_a\to0$). For both electrolytes the smeared-charge model reproduces the measured conductivity over the whole concentration interval, whereas the point-charge theory systematically deviates starting from $\approx10^{-2}$ M. This confirms that the non-local charge distributions captured by the Slater form factors are essential even for simple inorganic salts.

\begin{figure}
\centering
\includegraphics[width=0.75\linewidth]{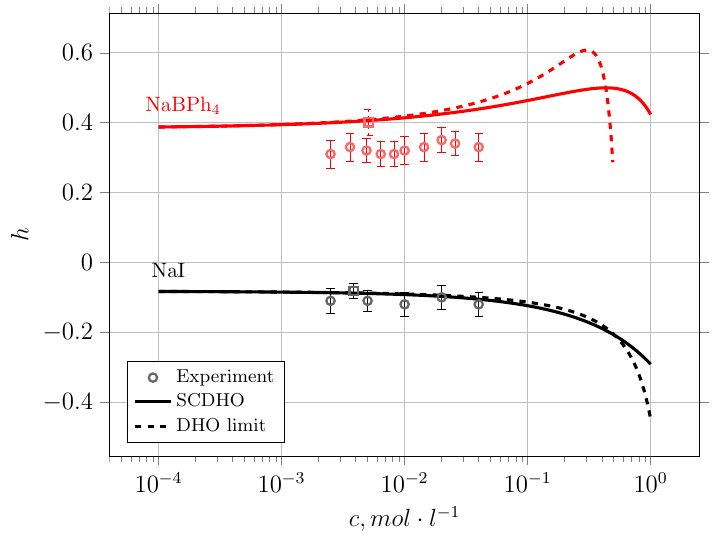}
\caption{Hall number $h$ of NaI and NaBPh$_4$ solutions versus molarity. Solid lines: smeared-charge SCDHO results; dashed lines: point-charge DHO limit. Experimental data are from Refs.~\cite{abbes1980effet,meton1976hall}. Error bars for the concentration series reproduce the uncertainties reported in Ref.~\cite{abbes1980effet}; those for the dilute measurements of Ref.~\cite{meton1976hall} are propagated from its Hall-voltage uncertainty.}
\label{h_vs_conc_salts}
\end{figure}

The concentration dependence of $h$ is compared with experiment~\cite{abbes1980effet,meton1976hall} in Fig.~\ref{h_vs_conc_salts}. The NaI Hall number remains negative and its magnitude increases with concentration. The signs and typical magnitudes of both salts are reproduced in the dilute range, although residual deviations from individual measurements remain.

For NaBPh$_4$, the Hall number is positive and nonmonotonic: the SCDHO curve reaches a broad maximum of $h\simeq0.50$ at $c\simeq0.4$ M and decreases at higher concentrations. This behavior results from competition between the reduction of the ordinary conductivity in the denominator of $h\propto\sigma^H/\sigma^2$ and the simultaneous suppression of the positive Hall conductivity by the interaction corrections. The maximum lies beyond the concentration range of the available Hall measurements and is a prediction of the model.

\begin{figure}
\centering
\includegraphics[width=\linewidth]{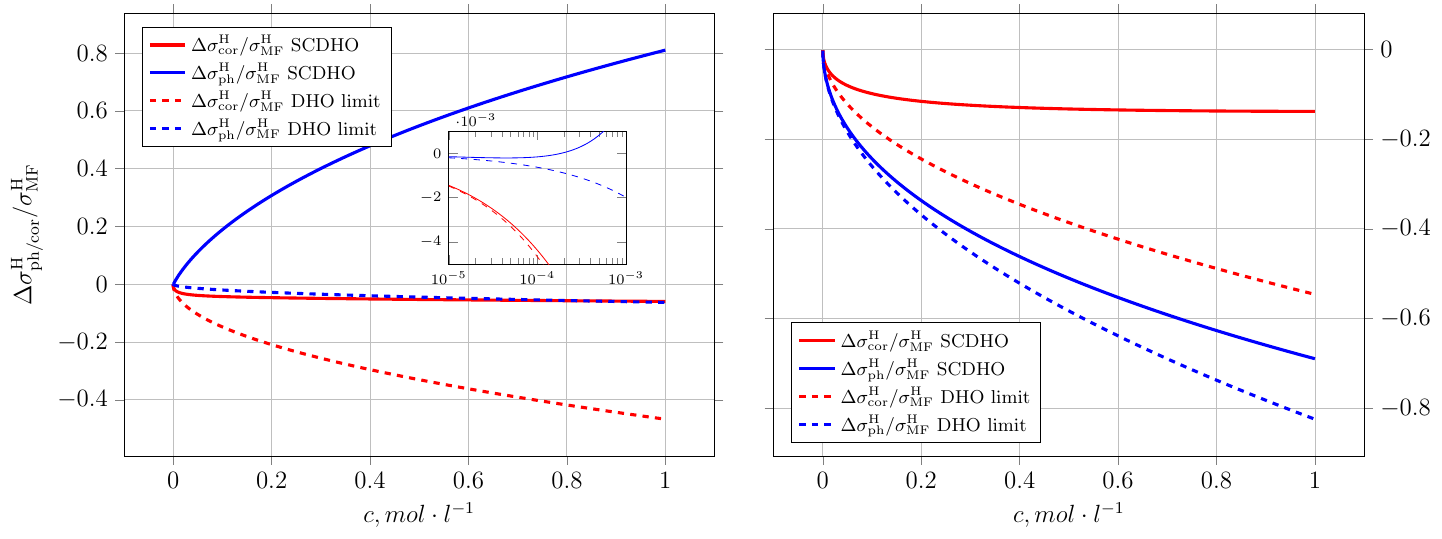}
\caption{Correlation and electrophoretic Hall-conductivity corrections for NaI (left) and NaBPh$_4$ (right), normalized by the corresponding mean-field Hall conductivity. Solid lines: SCDHO; dashed lines: point-charge DHO limit. Positive (negative) normalized values reinforce (oppose) the mean-field response.}
\label{contributions_salts}
\end{figure}

Figure~\ref{contributions_salts} separates the two interaction channels, normalized by the corresponding mean-field Hall conductivity. A positive normalized value reinforces the mean-field response, whereas a negative value opposes it; this convention is especially important when $\sigma_{\rm MF}^H<0$.

For NaI, the mean-field Hall conductivity is negative. The SCDHO correlation contribution opposes it throughout the displayed range. The electrophoretic contribution changes its orientation near $2\times10^{-4}$ M: at lower concentrations it opposes the mean-field term, while at higher concentrations it reinforces the negative mean-field response. The point-charge model does not exhibit this crossover. The distinct smearing lengths of Na$^+$ and I$^-$ affect the two electrophoretic vertices differently and are essential to this change of sign.

For NaBPh$_4$, the mean-field Hall conductivity is positive. Both interaction contributions oppose it over the displayed range, with the electrophoretic channel remaining the larger correction. Their effect on $h$ must be assessed together with the ordinary conductivity, rather than inferred from their signs alone.

\subsection{\texorpdfstring{Case 2: $\beta_+<\beta_-$, $m_+> m_-$ -- Acidic solutions.}{Case 2: Acidic solutions}}

We now turn to the case of two strong acids, HCl and HNO$_3$, which represent the opposite limit of mobility asymmetry: the proton has an exceptionally high mobility due to the Grotthuss mechanism, while the anion (Cl$^-$ or NO$_3^-$) moves by ordinary Stokes drift. Consequently, the mobility ratios are rather large: $m_{\text{H}^+}/m_{\text{NO}_3^-}\approx5$ and $m_{\text{H}^+}/m_{\text{Cl}^-}\approx4.6$. The dielectric friction is also far weaker on the proton; for the Hall comparison we adopt the experimental ionic Hall numbers for H$^+$ and Cl$^-$ from Ref.~\cite{meton1976hall} and restrict the analysis to HCl. The charge-smearing lengths for the anions were kept close to their previously determined values \cite{budkov2026conductivity}, while the smearing length for the proton was treated as a fitting parameter simultaneously for both acids (see Table \ref{input_parameters}).

\begin{figure}
    \centering
    \includegraphics[width=0.6\linewidth]{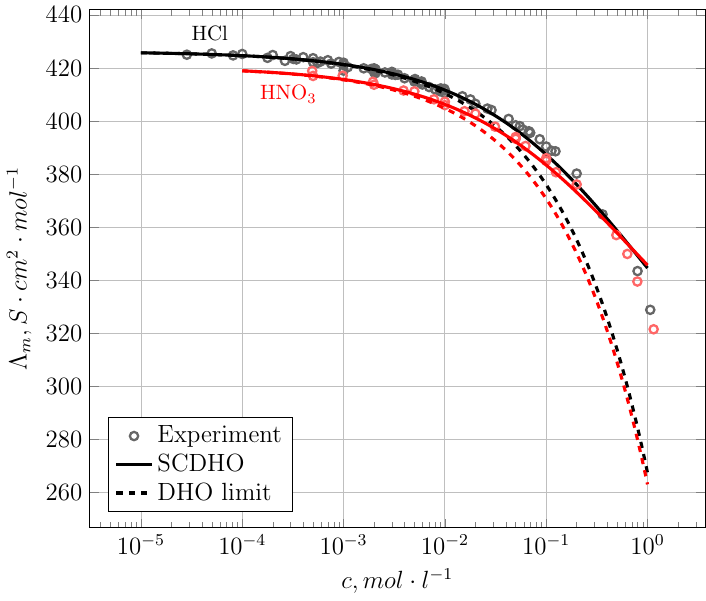}
    \caption{Molar electric conductivity of HCl and HNO$_3$ solutions versus molarity. Solid lines: SCDHO; dashed lines: point-charge DHO limit. Experimental data are from Ref.~\cite{Holze2016Electrochemistry}.}
    \label{electric_conductivity_acids}
\end{figure}

Figure \ref{electric_conductivity_acids} displays the molar conductivity $\Lambda_m$ of HCl and HNO$_3$ up to 1 M. The experimental data (symbols) are taken from the compilation \cite{Holze2016Electrochemistry}. The SCDHO model, which treats the proton's mobility as an experimental input through the conductivity at infinite dilution (i.e., no Stokes-Einstein assumption for H$^+$), reproduces the data within the studied concentration range very well when charge smearing is included. The curves corresponding to the point-charge approximation lie far below, illustrating once more the necessity of regularizing the short-range Coulomb interaction.

\begin{figure}
\centering
\includegraphics[width=0.75\linewidth]{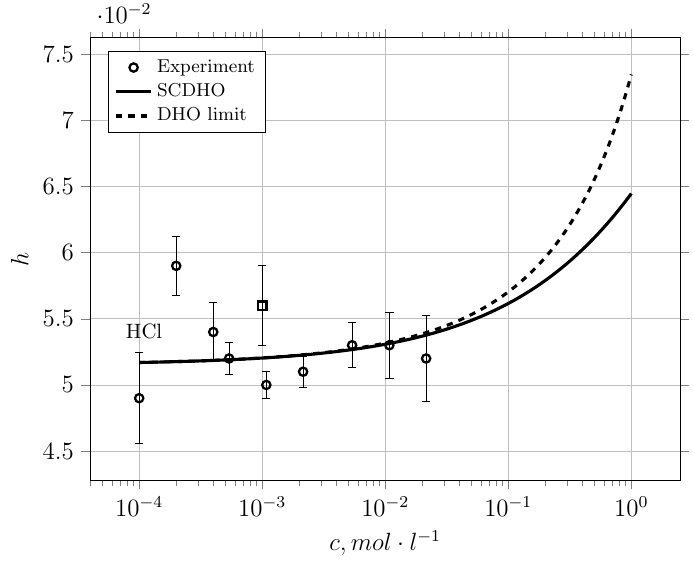}
\caption{Hall number $h$ of HCl solutions versus molarity. Solid line: smeared-charge SCDHO result; dashed line: point-charge DHO limit. Experimental data are from Refs.~\cite{gerard1990hall,meton1976hall}, with error bars propagated from the reported Hall-voltage uncertainties.}
\label{h_vs_conc_acids}
\end{figure}

The Hall analysis in Fig.~\ref{h_vs_conc_acids} is restricted to HCl. The HNO$_3$ Hall points shown in the published version~\cite{budkov2026hall} did not reproduce the values in the original experimental source~\cite{bellissent1971eperimental} and have been removed. The ordinary HNO$_3$ conductivity comparison in Fig.~\ref{electric_conductivity_acids} is unaffected.

For HCl, $h$ is positive, reflecting the dominant proton contribution despite its suppressed effective Lorentz response. The calculation captures the dilute Hall measurements~\cite{gerard1990hall,meton1976hall} within their scatter; the curves at higher concentrations remain predictions.

\begin{figure}
\centering
\includegraphics[width=0.75\linewidth]{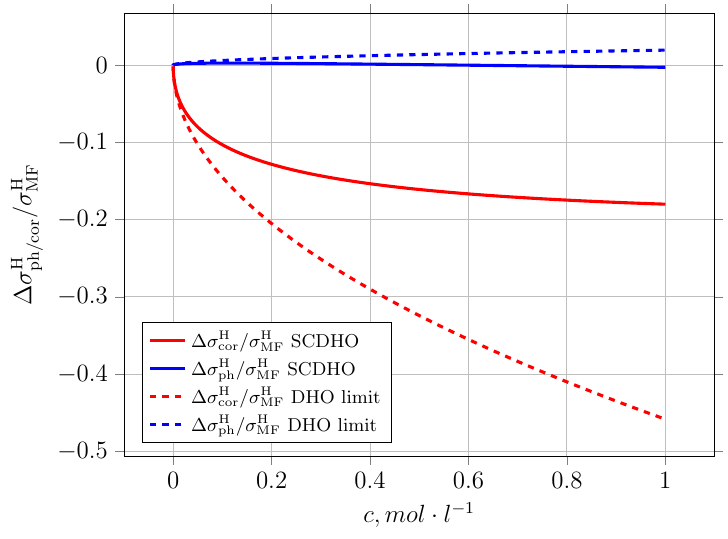}
\caption{Correlation and electrophoretic Hall-conductivity corrections for HCl, normalized by its mean-field Hall conductivity. Solid lines: SCDHO; dashed lines: point-charge DHO limit. Positive (negative) normalized values reinforce (oppose) the mean-field response.}
\label{contributions_acids}
\end{figure}

The decomposition for HCl in Fig.~\ref{contributions_acids} is dominated by the correlation contribution, which opposes the positive mean-field Hall response throughout the concentration range. The electrophoretic contribution is much smaller. It reinforces the mean-field term at low and moderate concentrations, crosses zero near $c\simeq0.6$ M, and opposes it at higher concentrations. The point-charge electrophoretic correction remains positive. Thus charge smearing changes the orientation of the smaller interaction channel without altering the dominant role of the correlation correction.

\subsection{\texorpdfstring{Case 3. $\beta_+>\beta_-$, $m_+<m_-$: Alkali hydroxide solutions.}{Case 3: Alkali hydroxide solutions}}

We now examine the three aqueous alkali hydroxide solutions LiOH, NaOH and KOH. In these systems the hydroxide ion OH$^-$ possesses an exceptionally high mobility due to a structural (proton-hole) diffusion mechanism analogous to that of the proton; its limiting ionic conductivity is second only to H$^+$ itself. The alkali cations, in contrast, move by ordinary Stokes drift, so the mobility ratio $m_{\text{OH}^-}/m_{\text{cation}}$ ranges from approximately 5 for Li$^+$ to 4 for Na$^+$ and 2.5 for K$^+$. The ionic Hall numbers adopted from Ref.~\cite{meton1976hall} (Table \ref{input_parameters}) are in the order Li$^+>$ Na$^+>$ K$^+$, while $\beta_{\text{OH}^-}$ is about 4--5 times smaller.

\begin{figure}
    \centering
    \includegraphics[width=0.6\linewidth]{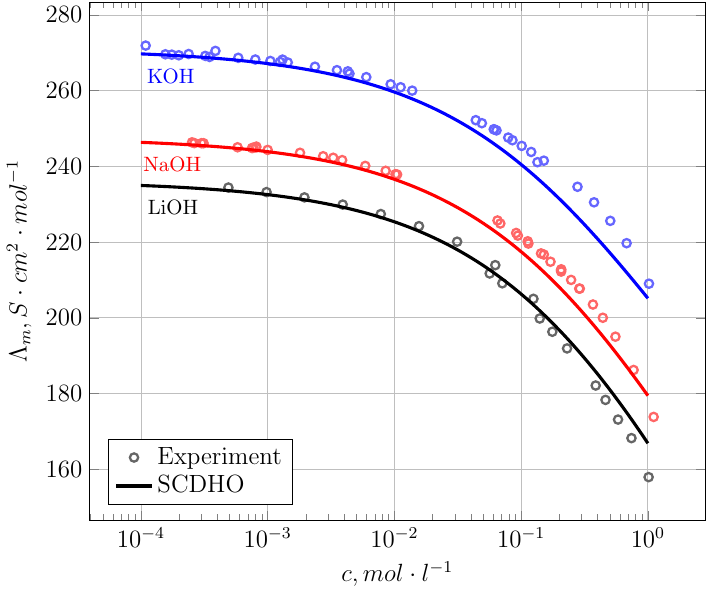}
    \caption{Molar electric conductivity of LiOH, NaOH, and KOH solutions versus molarity. Lines: SCDHO; symbols: experiment. Experimental data are from Ref.~\cite{Holze2016Electrochemistry}.}
    \label{electric_conductivity_hydroxides}
\end{figure}

Figure \ref{electric_conductivity_hydroxides} shows the molar conductivity $\Lambda_m$ as a function of concentration for LiOH, NaOH and KOH. The experimental data \cite{Holze2016Electrochemistry} cover the range from dilute solutions to about 1 M. The SCDHO predictions with the smeared ion charges (Table \ref{input_parameters}) reproduce the measurements over the entire concentration interval.

\begin{figure}
\centering
\includegraphics[width=0.75\linewidth]{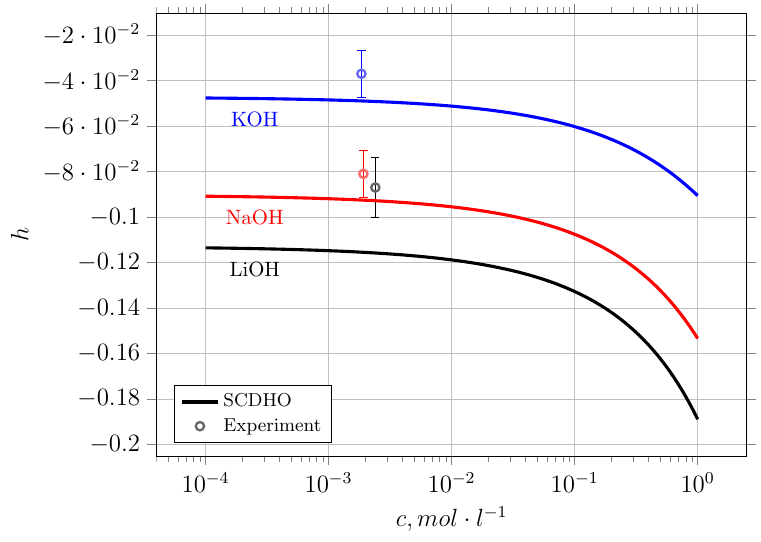}
\caption{SCDHO Hall numbers for LiOH, NaOH, and KOH solutions versus molarity. Lines show theory and symbols show dilute measurements from Ref.~\cite{meton1976hall}; error bars are propagated from the reported Hall-voltage uncertainty $\Delta V_{\rm H}\simeq20$ nV.}
\label{h_vs_conc_hydroxides}
\end{figure}

The Hall number is negative for all three hydroxides (Fig.~\ref{h_vs_conc_hydroxides}), with $|h_{\rm LiOH}|>|h_{\rm NaOH}|>|h_{\rm KOH}|$, and its magnitude increases with concentration. The dilute experimental points have the same sign and order of magnitude. The ordering is already largely determined by the ion mobilities and Hall coefficients in the mean-field term, while the interaction corrections control the concentration dependence.

\begin{figure}
\centering
\includegraphics[width=0.9\linewidth]{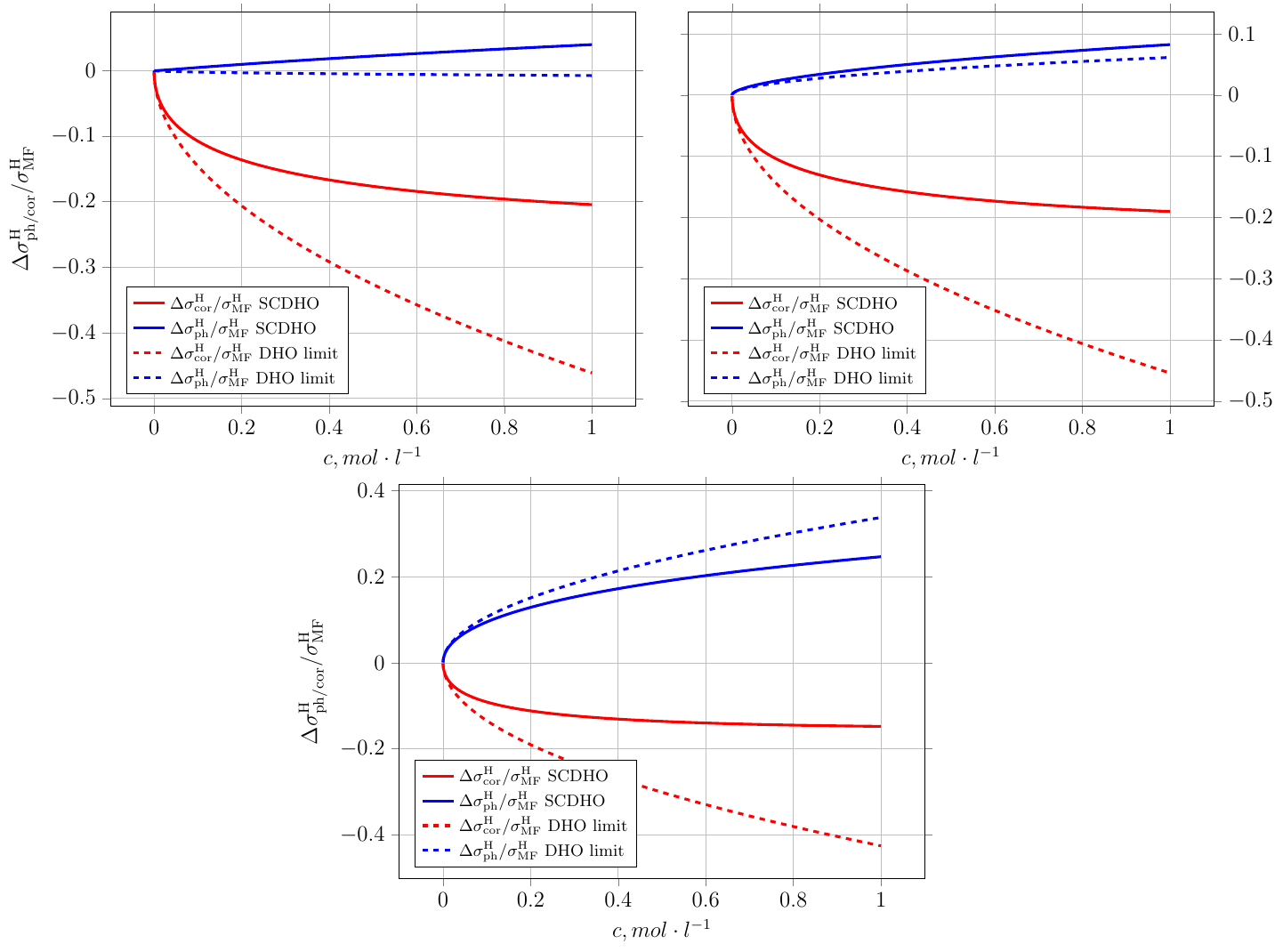}
\caption{Correlation and electrophoretic Hall-conductivity corrections for LiOH (upper left), NaOH (upper right), and KOH (lower panel), normalized by the corresponding mean-field Hall conductivity. Solid lines: SCDHO; dashed lines: point-charge DHO limit. Positive (negative) normalized values reinforce (oppose) the mean-field response.}
\label{contributions_hydroxides}
\end{figure}

Figure~\ref{contributions_hydroxides} shows the redistribution of the interaction channels along the LiOH--NaOH--KOH series. The mean-field Hall conductivity is negative in each case. The SCDHO correlation correction opposes it, while the electrophoretic correction becomes increasingly important from LiOH to KOH and reinforces the mean-field response at higher concentrations. The correlation channel dominates for LiOH; for KOH, the two corrections are nearly comparable in magnitude. Their partial cancellation is a concentration-dependent correction superimposed on the dilute-limit hierarchy.

\subsection{\texorpdfstring{Case 4. $\beta_+\approx\beta_-$, $m_+<m_-$: Ionic liquid solutions.}{Case 4: Aqueous imidazolium-salt solutions}}

As a final application and test of the model, we consider two aqueous solutions of imidazolium-based ionic liquids, 1-hexyl-3-methylimidazolium bromide (HmimBr) and chloride (HmimCl). In these systems the small anion is the more mobile species ($m_{\text{Br}^-}\approx m_{\text{Cl}^-}>m_{\text{Hmim}^+}$) with the mobility ratio $m_{\text{anion}}/m_{\text{Hmim}^+}\approx2.5$, leading to a negative mean-field Hall conductivity. The ionic Hall number for Hmim$^+$ is estimated via Eq.~(\ref{beta_a_det}), while for anions they are taken from Ref.~\cite{meton1976hall}; the charge-smearing lengths were fixed by fits of the Ohmic conductivity. No experimental Hall data are available for these salts; the results presented here are genuine predictions of the SCDHO theory. We emphasize that the present treatment is formulated for aqueous electrolyte solutions with an incompressible solvent background. In ionic-liquid-rich mixtures, where the mass and volume fractions of the ionic liquid become comparable to that of water, total momentum conservation and volume-flow constraints may require a more complete multicomponent hydrodynamic formulation.

\begin{figure}
    \centering
    \includegraphics[width=0.6\linewidth]{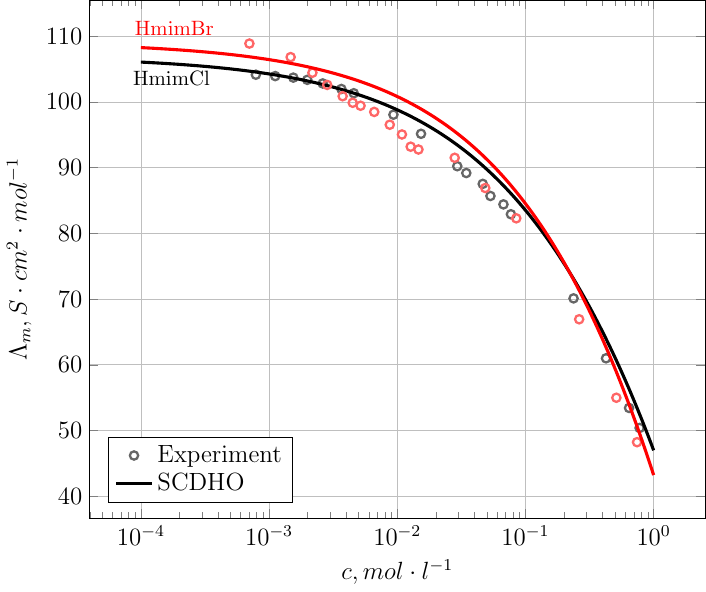}
    \caption{Molar electric conductivity of aqueous HmimCl and HmimBr solutions versus molarity. Lines: SCDHO; symbols: experiment. Experimental data are from Ref.~\cite{wang2023experimental}.}
    \label{electric_conductivity_ILs}
\end{figure}

Figure \ref{electric_conductivity_ILs} compares the experimental molar conductivity \cite{wang2023experimental} with the SCDHO predictions for HmimBr and HmimCl. The smeared-charge model reproduces the measured data over the full concentration range up to 1 M.

\begin{figure}
\centering
\includegraphics[width=0.75\linewidth]{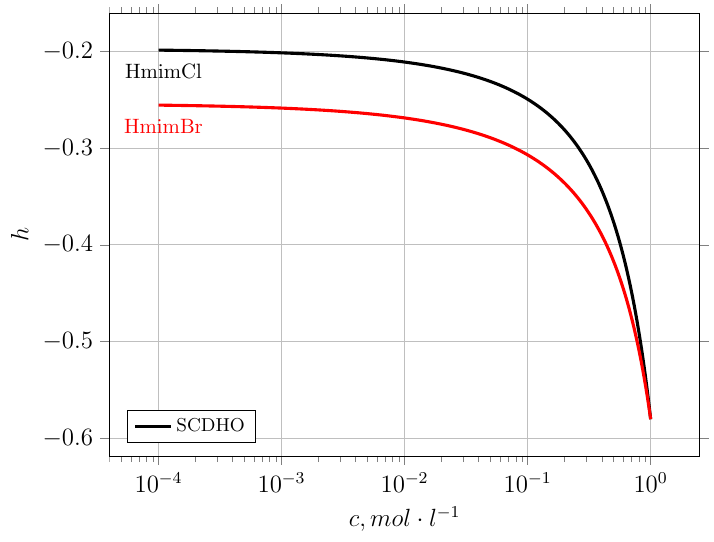}
\caption{Predicted SCDHO Hall numbers for aqueous HmimCl and HmimBr solutions versus molarity. No experimental Hall data are available for these systems.}
\label{h_vs_conc_ILs}
\end{figure}

The predicted Hall number is negative for both aqueous imidazolium salts (Fig.~\ref{h_vs_conc_ILs}). Its magnitude increases with concentration and is larger for HmimBr than for HmimCl over most of the range; the curves nearly meet at 1 M. No experimental Hall data are available for these systems.

\begin{figure}
\centering
\includegraphics[width=\linewidth]{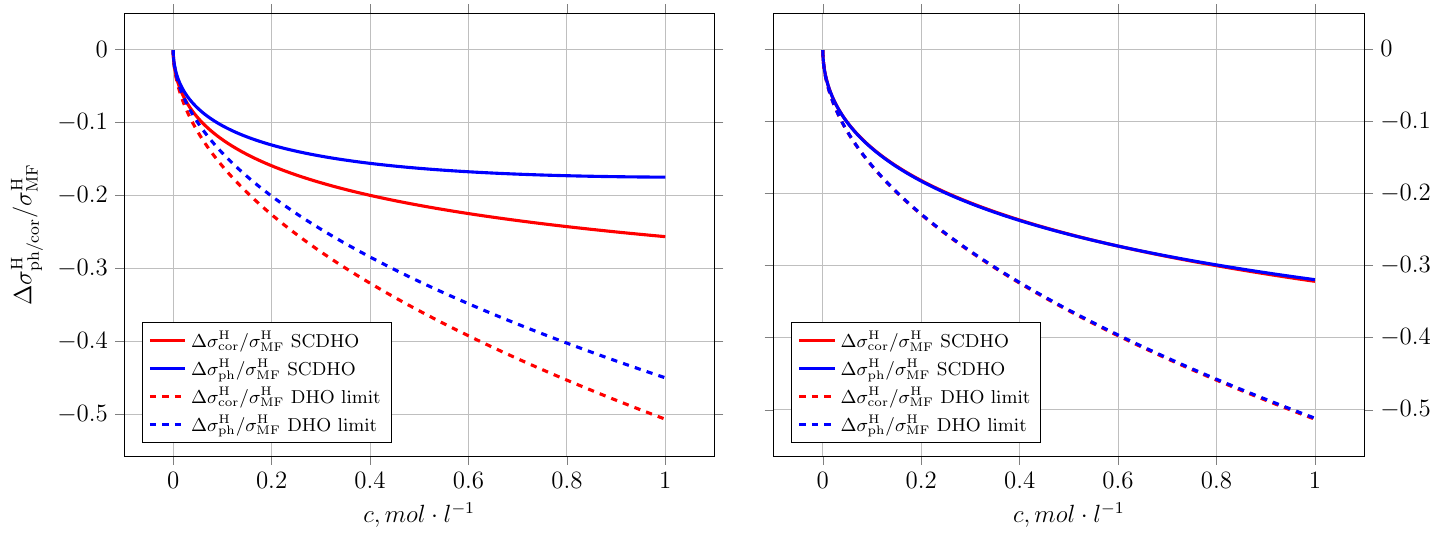}
\caption{Correlation and electrophoretic Hall-conductivity corrections for HmimCl (left) and HmimBr (right), normalized by the corresponding mean-field Hall conductivity. Solid lines: SCDHO; dashed lines: point-charge DHO limit. Negative normalized values oppose the negative mean-field response.}
\label{contributions_ILs}
\end{figure}

Both interaction corrections in Fig.~\ref{contributions_ILs} have negative values when normalized by the negative mean-field Hall conductivity and therefore oppose the mean-field response. For HmimCl, the correlation correction is moderately larger in magnitude than the electrophoretic one; for HmimBr, the two nearly coincide over a broad concentration range. Charge smearing reduces their magnitudes relative to the point-charge limit while preserving their comparable scale.

\section{Discussion and Concluding Remarks}

We have extended the self-consistent Debye--H\"uckel--Onsager framework~\cite{budkov2026conductivity} to electrolyte Hall transport. The Hall current contains a mean-field term and two interaction channels. The correlation channel combines the magnetic distortion of the ionic atmosphere with the Lorentz rotation of the ordinary relaxation current. The electrophoretic channel combines the vortical solvent response with the Lorentz rotation of the ordinary back-flow. Retaining all four interaction vertices is necessary for a consistent weak-field Hall current.

Within the smeared-charge RPA, the interaction corrections reduce to one-dimensional wave-number integrals, with closed analytical point-charge limits. When hydrodynamics is suppressed, the point-charge relaxation correction agrees exactly with odd stochastic density functional theory~\cite{avni2026hall} and the relaxation part of Friedman's result~\cite{friedman1965calculation}. The comparison does not establish equivalence of the full hydrodynamic theories.

The non-local charge distribution can change both the magnitude and the sign of individual Hall corrections. In NaI, the SCDHO electrophoretic correction changes from opposing to reinforcing the negative mean-field response near $2\times10^{-4}$ M, while the correlation correction continues to oppose it. For NaBPh$_4$, both interaction terms oppose the positive mean-field Hall conductivity, yet the competing reduction of the ordinary conductivity produces a broad maximum of $h$ near 0.4 M. For HCl, the smaller electrophoretic correction crosses zero near 0.6 M and the correlation channel remains dominant. Along the hydroxide series, correlation effects dominate for LiOH and the two channels become nearly comparable for KOH. They are also comparable for the aqueous imidazolium salts.

The ion-specific dielectric friction coefficients $\beta_a$ are essential for quantitative comparison with experiment. For ions transported by ordinary hydrodynamic drift, the electrohydrodynamic estimates of Hubbard and Wolynes and of Kroh and Felderhof~\cite{hubbard1981electrohydrodynamic,kroh1987hall} provides a natural physical estimate of the reduction of the Lorentz force. For protons and hydroxide ions, however, the Grotthuss hopping mechanism makes the effective Hall response much smaller than that of ordinary Brownian ions, and the corresponding values of $\beta_{\mathrm{H}^+}$ and $\beta_{\mathrm{OH}^-}$ must be taken from experiment. This semi-empirical input allows one to describe acids and hydroxides within the same theoretical framework as simple salts, but it also indicates that a fully microscopic theory of the Hall response of structure-diffusing ions remains an open problem.

For the 22-system dilute benchmark, the mean absolute relative deviations are about 17\% and 12 systems lie within the reported experimental uncertainty. The finite-concentration Hall measurements are much more limited than the ordinary-conductivity data, so the predicted behavior at moderate and high concentrations requires further experimental tests. In particular, the NaBPh$_4$ maximum and the HCl electrophoretic crossover are model predictions outside the available Hall-data range. The incorrectly reproduced HNO$_3$ Hall points have been removed; the HNO$_3$ ordinary-conductivity comparison is retained.

Residual discrepancies can arise from the RPA closure, neglect of ion pairing and specific short-range interactions, the continuum solvent description, and the omission of the feedback of electrophoretic flow on steady-state pair correlations. No new fit of mobilities, Hall coefficients, or charge-smearing lengths has been introduced by the corrections. The predictions therefore retain their sensitivity to the original input parameters.

Several extensions of the present theory are therefore natural. First, the linear-in-$\mathbf B$ approximation can be generalized to stronger magnetic fields, where higher-order magnetotransport effects and field-induced distortions of the ionic atmosphere may become relevant. Second, short-range repulsion, ion association, and specific solvation effects can be incorporated by replacing the random-phase approximation with more accurate integral-equation or classical density-functional descriptions, in the spirit of MSA-based and related approaches to concentrated electrolyte transport~\cite{bernard1992conductance,bernard2023analytical}. Third, the theory can be generalized to multicomponent electrolytes, non-aqueous solvents, ionic-liquid mixtures, deep eutectic solvents, and battery electrolytes, where the conservation of total momentum and volume flow may require a more complete hydrodynamic formulation. Finally, coupling the present framework to stochastic density functional theory~\cite{demery2016conductivity,avni2020charge,avni2022conductance} would make it possible to address fluctuation-driven and time-dependent Hall responses.

Overall, the SCDHO extension developed here provides a compact and physically interpretable theory of electrolyte Hall conductivity in terms of a limited set of parameters: ionic mobilities, charges, charge-smearing lengths, and ion-specific Hall coefficients. The theory not only rationalizes the available experimental data but also gives experimentally testable predictions for systems where Hall measurements have not yet been performed. It therefore provides a useful basis for future theoretical and experimental studies of magnetotransport in liquid electrolytes, with possible applications in nanofluidics, iontronics, and magneto-electrochemistry.

\textbf{Data availability statement.} {\sl The data supporting the findings of this study are available in the article.}

\textbf{Acknowledgements.}
The calculations were performed using computational resources from HPC facilities at NRU HSE. The article was prepared within the framework of the project 'Mirror Laboratories' HSE University. We thank Yael Avni for drawing our attention to the discrepancy in the dilute point-charge limit and for a useful comparison with the independent odd-SDFT calculation.

\appendix

\section{\texorpdfstring{Derivation of Eq.~\eqref{basic}}{Derivation of the pair-transport equation}}
\label{app:derivation_basic}

In this Appendix we give the intermediate steps leading from the two-particle continuity equation to Eq.~\eqref{basic}.  The purpose is to make explicit how the Ornstein--Zernike relation enters the linearized chemical-potential term and how the magnetic part of the drift force produces the vorticity terms on the left-hand side of Eq.~\eqref{basic}.

We start from the two-particle stationary continuity equation,
\begin{equation}
\label{app_continuity}
\nabla_a\cdot\mathbf j_{ab}(\mathbf r_a,\mathbf r_b)
+
\nabla_b\cdot\mathbf j_{ba}(\mathbf r_b,\mathbf r_a)=0,
\end{equation}
where
\begin{equation}
\mathbf j_{ab}=g_{ab}\mathbf v_{ab},
\qquad
\mathbf v_{ab}=m_a\mathbf f_a-m_a\nabla_a\mu_a+
\mathbf u_b .
\end{equation}
Substitution gives
\begin{multline}
\label{app_full_cont}
\nabla_a\cdot\left[g_{ab}\left(m_a\mathbf f_a+
\mathbf u_b\right)\right]
-
\nabla_a\cdot\left[g_{ab}m_a\nabla_a\mu_a\right]\\
+
\nabla_b\cdot\left[g_{ba}\left(m_b\mathbf f_b+
\mathbf u_a\right)\right]
-
\nabla_b\cdot\left[g_{ba}m_b\nabla_b\mu_b\right]=0 .
\end{multline}
The equilibrium part of this equation is identically satisfied.  We therefore keep only the terms linear in the external fields and in the nonequilibrium distortion of the pair correlations.  In the diffusive terms it is sufficient to put $g_{ab}\simeq 1$, whereas in the drift terms the product of the external force with the equilibrium inhomogeneity of the ionic atmosphere must be retained.  Thus
\begin{multline}
\label{app_linear_cont}
-m_a\nabla_a^2\delta\mu_{a}^{\rm neq}(\mathbf r_a,\mathbf r_b)
-m_b\nabla_b^2\delta\mu_{b}^{\rm neq}(\mathbf r_b,\mathbf r_a)\\
+
\left(m_a\mathbf f_a+\mathbf u_b\right)\cdot
\nabla_a h_{ab}(\mathbf r_a,\mathbf r_b)
+
\left(m_b\mathbf f_b+\mathbf u_a\right)\cdot
\nabla_b h_{ba}(\mathbf r_b,\mathbf r_a)\\
+
\nabla_a\cdot\left(m_a\mathbf f_a+\mathbf u_b\right)
+
\nabla_b\cdot\left(m_b\mathbf f_b+\mathbf u_a\right)=0 .
\end{multline}
Here $\delta\mu_a^{\rm neq}$ denotes the part of the local chemical potential produced by the nonequilibrium distortion $\omega_{ab}$ of the pair correlation function.

Let us now evaluate this nonequilibrium part.  From Eq.~\eqref{chem_pot_lin} and the decomposition
\begin{equation}
 h_{ab}(\mathbf r_a,\mathbf r_b)=
 \bar h_{ab}(\mathbf r_a-\mathbf r_b)+
 \omega_{ab}(\mathbf r_a,\mathbf r_b),
\end{equation}
one obtains, for the part linear in $\omega$,
\begin{equation}
\label{app_mu_omega_raw}
\frac{\delta\mu_a^{\rm neq}(\mathbf r_a,\mathbf r_b)}{k_B T}
=
\omega_{ab}(\mathbf r_a,\mathbf r_b)
-
\sum_l\int d^3\mathbf r'\,
 c_{al}(\mathbf r_a-\mathbf r')\,
 \bar c_l\omega_{lb}(\mathbf r',\mathbf r_b).
\end{equation}
The fixed-particle delta contribution and the equilibrium part have been removed by subtracting the equilibrium relation.  Equivalently, the cancellation of the equilibrium terms follows from the Ornstein--Zernike equation
\begin{equation}
\bar h_{ab}(\mathbf r_a-\mathbf r_b)
=
 c_{ab}(\mathbf r_a-\mathbf r_b)
+
\sum_l\int d^3\mathbf r'\,
 c_{al}(\mathbf r_a-\mathbf r')\bar c_l
 \bar h_{lb}(\mathbf r'-\mathbf r_b),
\end{equation}
which is precisely Eq.~\eqref{OZ}.  It is therefore natural to introduce
\begin{equation}
\label{app_hat_omega}
\hat\omega_{ab}(\mathbf r_a,\mathbf r_b)=
\omega_{ab}(\mathbf r_a,\mathbf r_b)
-
\sum_l\int d^3\mathbf r'\,
 c_{al}(\mathbf r_a-\mathbf r')\bar c_l
 \omega_{lb}(\mathbf r',\mathbf r_b),
\end{equation}
so that
\begin{equation}
\label{app_mu_hat}
\delta\mu_a^{\rm neq}(\mathbf r_a,\mathbf r_b)=k_B T\,
\hat\omega_{ab}(\mathbf r_a,\mathbf r_b),
\qquad
\delta\mu_b^{\rm neq}(\mathbf r_b,\mathbf r_a)=k_B T\,
\hat\omega_{ba}(\mathbf r_b,\mathbf r_a).
\end{equation}
Substitution of Eq.~\eqref{app_mu_hat} into Eq.~\eqref{app_linear_cont} gives
\begin{multline}
\label{app_before_div}
 m_a k_B T\nabla_a^2\hat\omega_{ab}(\mathbf r_a,\mathbf r_b)
+
 m_b k_B T\nabla_b^2\hat\omega_{ba}(\mathbf r_b,\mathbf r_a)\\
=
\left(m_a\mathbf f_a+\mathbf u_b\right)\cdot
\nabla_a h_{ab}(\mathbf r_a,\mathbf r_b)
+
\left(m_b\mathbf f_b+\mathbf u_a\right)\cdot
\nabla_b h_{ba}(\mathbf r_b,\mathbf r_a)\\
+
\nabla_a\cdot\left(m_a\mathbf f_a+\mathbf u_b\right)
+
\nabla_b\cdot\left(m_b\mathbf f_b+\mathbf u_a\right).
\end{multline}

It remains to transform the divergence terms.  Because the solvent is incompressible,
\begin{equation}
\nabla_a\cdot\mathbf u_b=0,
\qquad
\nabla_b\cdot\mathbf u_a=0 .
\end{equation}
The electric part of $\mathbf f_a$ is spatially homogeneous and therefore has zero divergence.  Inside the Lorentz force, the complete electric-field-induced velocity must be used. The homogeneous drift has zero divergence, and the relaxation part contributes no divergence because, for a uniform magnetic field,
\begin{equation}
\nabla\cdot[(\nabla\delta\mu^E)\times\mathbf B]
=\mathbf B\cdot[\nabla\times\nabla\delta\mu^E]=0.
\end{equation}
Consequently, only the solvent velocity contributes to the divergence at this order, although both rotated interaction velocities remain necessary in the average current:
\begin{align}
\nabla_a\cdot\mathbf f_a
&\simeq
\beta_a q_a\nabla_a\cdot\left[\mathbf u_b(\mathbf r_a,\mathbf r_b)\times\mathbf B\right]
=
\beta_a q_a\mathbf B\cdot
\left[\nabla_a\times\mathbf u_b(\mathbf r_a,\mathbf r_b)\right],\\
\nabla_b\cdot\mathbf f_b
&\simeq
\beta_b q_b\nabla_b\cdot\left[\mathbf u_a(\mathbf r_b,\mathbf r_a)\times\mathbf B\right]
=
\beta_b q_b\mathbf B\cdot
\left[\nabla_b\times\mathbf u_a(\mathbf r_b,\mathbf r_a)\right].
\end{align}
Here we used the vector identity
\begin{equation}
\nabla\cdot(\mathbf u\times\mathbf B)=
\mathbf B\cdot(\nabla\times\mathbf u)-
\mathbf u\cdot(\nabla\times\mathbf B),
\end{equation}
and the external magnetic field is homogeneous, $\nabla\times\mathbf B=0$.
Moving these divergence terms to the left-hand side of Eq.~\eqref{app_before_div}, one obtains
\begin{multline}
 m_a k_B T \nabla_a^2 \hat{\omega}_{ab}(\mathbf r_a,\mathbf r_b)
 +m_b k_B T \nabla_b^2 \hat{\omega}_{ba}(\mathbf r_b,\mathbf r_a)\\
 -\beta_a m_a q_a\mathbf B\cdot
 [\nabla_a\times\mathbf u_b(\mathbf r_a,\mathbf r_b)]
 -\beta_b m_b q_b\mathbf B\cdot
 [\nabla_b\times\mathbf u_a(\mathbf r_b,\mathbf r_a)]\\
=
\left(m_a\mathbf f_a(\mathbf r_a,\mathbf r_b)+\mathbf u_b(\mathbf r_a,\mathbf r_b)\right)\cdot
\nabla_a h_{ab}(\mathbf r_a,\mathbf r_b)\\
+
\left(m_b\mathbf f_b(\mathbf r_b,\mathbf r_a)+\mathbf u_a(\mathbf r_b,\mathbf r_a)\right)\cdot
\nabla_b h_{ba}(\mathbf r_b,\mathbf r_a),
\end{multline}
which is Eq.~\eqref{basic}, up to the notational convention $a\leftrightarrow b$ used in the main text.  This derivation also shows explicitly why the Ornstein--Zernike relation is needed: it converts the linearized excess-chemical-potential contribution into the compact operator $\hat\omega_{ab}$ acting on the nonequilibrium distortion of the ionic atmosphere.

\bibliography{lit}

\end{document}